**Ridge Regression Estimated Linear Probability Model Predictions of *O*-glycosylation in Proteins with Structural and Sequence Data**

Rajaram Gana[†, ✉]     and     Sona Vasudevan[†, ✉]

[†] Department of Biochemistry and Molecular & Cellular Biology

Georgetown University, Washington D.C., USA

[✉] Corresponding author (contact emails: sv67@georgetown.edu and rg476@georgetown.edu)

# Abstract

The likelihood of *O*-GlcNAc glycosylation in human proteins is predicted using the ridge regression estimated linear probability model (LPM). To achieve this, sequences from three similar post-translational modifications (PTMs) of proteins occurring at, or very near, the *S/T*-site are analyzed: *N*-glycosylation, *O*-mucin type (*O*-GalNAc) glycosylation, and phosphorylation. Results found include: **1)** The consensus *composite* sequon for *O*-glycosylation is: ~ *(W–S/T–W)*, where "~" denotes the "not" operator. Thus, ~*W – S/T – W*, *W – S/T – ~W*, or ~*W – S/T – ~W* are necessary for *O*-glycosylation. **2)** The consensus sequon for phosphorylation is ~ *(W–S/T/Y/H–W);* although *W–S/T/Y/H–W* is not an absolute inhibitor of phosphorylation. **3)** For LPM estimation, *N*-glycosylated sequences are found to be good approximations to non-*O*-glycosylatable sequences; although *N – ~P – S/T* is not an absolute inhibitor of *O*-glycosylation. **4)** The selective positioning of an amino acid along the sequence, differentiates the PTMs of proteins. **5)** Some *N*-glycosylated sequences are also phosphorylated at the *S/T*-site in the *N – ~P – S/T* sequon. **6)** ASA values for *N*-glycosylated sequences are stochastically larger than those for *O*-GlcNAc glycosylated sequences. **7)** Structural attributes (beta turn II, II´, helix, beta bridges, beta hairpin, and the phi angle) are significant LPM predictors of *O*-GlcNAc glycosylation. The LPM with sequence *and* structural data as explanatory variables yields a Kolmogorov-Smirnov (KS) statistic value of 99%. **8)** With only sequence data, the KS statistic erodes to 80%, underscoring the germaneness of structural information, which is sparse on *O*-glycosylated sequences. With 50% as the cutoff probability for predicting *O*-GlcNAc glycosylation, this LPM mispredicts 21% of out-of-sample *O*-GlcNAc glycosylated sequences as not being glycosylated. The 95% confidence interval around this mispredictions rate is 16% to 26%.

Acknowledgments: We thank Swagata Naha, who was a graduate student at Georgetown University, for collecting data on some of the sequences. We thank Zhang-Zhi Hu for providing the data mined for the creation of dbOGAP during his tenure, as a faculty member, at Georgetown University. We thank Elliott Crooke, Professor and Chair, Department of Biochemistry and Molecular & Cellular Biology, and Senior Associate Dean, Faculty and Academic Affairs, Georgetown University, for his support and encouragement. We thank Dr. Sarma Dittakavi, Professor Emeritus, Laboratory Medicine and Pathobiology, University of Toronto, for reading the manuscript, providing valuable comments, and for the many stimulating discussions that we have had together.

*This paper was originally uploaded to arXiv on May 24, 2019. This present version, uploaded on Feb. 16, 2019, is essentially the same as the original version, but includes a considerably expanded "Concluding Remarks" section for clarificatory purposes.*

## Introduction

The process of post-translational modification (PTM) of proteins has allowed cells to vastly increase the otherwise normal array of functions that proteins carry out. PTMs are essential modifications that occur in higher eukaryotes. Glycosylation is one type of modification that chemically alters proteins by the enzymatic or non-enzymatic addition of glycans. There are two main types of glycosylation present: *N*-linked and *O*-linked [1]. There are two types of *O*-linked glycosylation in cells: mucin-type *O*-linked glycosylation (*O*-GalNAc); and several types of non-mucin *O*-glycans, which include α-linked *O*-fucose, β-linked *O*-xylose, α-linked *O*-mannose, β-linked *O*-GlcNAc (*N*-Acetyl Glucosamine), α- or β-linked *O*-galactose, α- or β-linked *O*-glucose glycans, and *O*-GlcNAcylation (or *O*-GlcNAc). In addition, *C*-linked, *S*-linked and non-enzymatic glycation processes exist [2]. The consensus sequon for *N*-linked glycosylation is $N - X - S/T$, where $X$ is not proline. Two consensus sequons for *C*-linked glycosylation are: $W - X - X - W$ and $W - S/T - X - C$ [3]. No consensus sequon for *O*-glycosylation is known thus far.

*O*-GlcNAc modification is the attachment of β-*N*-Acetyl Glucosamine (GlcNAc) to protein serine (*S*) or threonine (*T*) amino acid residues *via* the *O*-linked glycosylation process [4]. *O*-GlcNAcylation occurs primarily in nucleocytoplasmic proteins [5, 6]. *O*-GlcNAcylation mechanism targets numerous transcription factors, tumor suppressors, kinases, phosphatases, and histone-modifying proteins [6-9]. Furthermore, the monoaddition of GlcNAc is suggested to be the most significant glycosylation process in the regulation of metazoan cytosolic and nuclear proteins [10, 11]. Often, the *S/T* residues, which are targeted by *O*-GlcNAcylation, are also targets of kinases and phosphatases. Hence, the consensus sequon for *O*-linked glycosylation can also be the one for phosphorylation. These residues are both dynamic and inducing in nature [12]. Thus, to summarize, *O*-GlcNAcylation plays a major role in modulating kinase-induced signal transduction pathways [13], supporting the notion that *O*-GlcNAcylation has an important biological role.

The source of GlcNAc for *O*-GlcNAcylation in eukaryotic cells is UDP-*N*-Acetyl Glucosamine [14], UDP-GlcNAc, obtained from glucose in the hexosamine biosynthetic pathway (HBP). The attachment process of *O*-GlcNAcylation is catalyzed *via* *O*-linked β-*N*-Acetyl Glucosamine transferase [15, 16] or *O*-GlcNAc transferase and is removed by *O*-linked β-*N*-acetyl- glucosaminidase or *O*-GlcNAcase [16]. Currently, *O*-GlcNAc transferase is the only known enzyme in mammals responsible for this addition of GlcNAc [17]. *O*-GlcNAc transferase is found in the nucleus, mitochondria, and cytoplasm of cells.

*O*-GlcNAc transferase has important roles in many biological processes and diseases. These include type 2 diabetes and its complications, such as blindness, renal failure, and nerve damage; diabetes-accelerated atherosclerosis; dyslipidemia; cardiovascular diseases; aging and cancer [18-23]. Additionally, *O*-GlcNAc transferase is associated with tauopathies, including Alzheimer's disease, dementia, and Parkinson's disease [24, 25]. Recently, the role of glycosylation in neuroblastoma has taken center stage [26].

The common underlying cause of such diseases is due to the over- or under- active process of *O*-GlcNAcylation. With the complex metabolic processes in *O*-GlcNAcylation and its critical role in the regulation of cellular processes, it is not unusual to consider abnormal *O*-GlcNAcylation activity as contributing to various diseased states. This concept suggests that an in-depth understanding of sequence and structural parameters that drive *O*-GlcNAcylation may shed some light on *O*-linked glycosylation.

The goal of this paper is to use the ridge regression (RR) estimated linear probability model (LPM) to predict the likelihood of *O*-GlcNAc glycosylation in human proteins [27-33]. A detailed description of the statistical methodology used herein is in a predecessor paper [34]. The LPM models a binary outcome: *O*-GlcNAc glycosylated or not. So, the LPM estimation dataset requires two data subsets: *O*-GlcNAc glycosylated sequences and sequences that cannot be *O*-GlcNAc glycosylated. Experimentally one can determine if a protein is *O*-glycosylated. Thus, getting data on *O*-glycosylated proteins is straightforward. Getting data on proteins that are highly unlikely (or impossible) to be *O*-glycosylated is quite challenging, because a consensus sequon for *O*-glycosylation is needed. To understand whether a true, or even approximate, consensus sequon for *O*-glycosylation exists, a family of similar sequences with PTMs of proteins occurring at the *S/T*-site is collected. This family includes sequences on the following PTMs of proteins: *N*-glycosylation, *O*-GalNAc glycosylation, and phosphorylation. The rest of this paper documents the analyses done with all of this collected data.

## Materials and Methods

### *Data Collection*

a.  Data collected on the glycosylated sequences used for analysis herein is stored in an Excel spreadsheet: *glycos_public.xlsx* (which is available upon request). This data, and key operations performed on it, are described in Table 1. Data is collected from two resources (dbOGAP [35] and PhosphoSitePlus [36]) as described in Table 1.

b.  Structural data is obtained from the Protein Data Bank (PDB) [37], and the PDB-ID codes used are provided in *glycos_public.xlsx* in the column labeled *PDBID*. In dbOGAP, the total number of UniProt sequences is 376; and, of these, 39 have structural information. Several unique structures exist for each of the 39 sequences. This results in the 39 sequences having a total of 998 structures with unique PDB-IDs. Each structure is unique in terms of different ligands/drugs bound to them. Although there is redundancy in this data from a sequence standpoint, these 39 sequences have different structural attributes as defined by phi-psi angles and ASA, which are important structural variables. In order to make sure that these are real structural differences, similar structures were superposed to calculate a RMSD (root-mean-squared-deviation). It is confirmed that the resolution of the structures is high enough to believe in the subtle conformational changes captured in the data. The 10 residues to the left of the *S/T*-site, and the 10 residues to the right of the *S/T*-site, are superposed to see if the *S/T*-site displays slightly different conformations. An example is presented in Figure 1. This example supports the rationale behind using all of the redundant sequences for regression model building. It is not surprising that the *S*-site, in this example, is flexible enough to accommodate its role in a multitude of cellular processes. Residues that have maximum deviation (0.5-1.0Å) are close to the *S*-site. Residues farther out superimpose with less than 0.1Å RMSD.

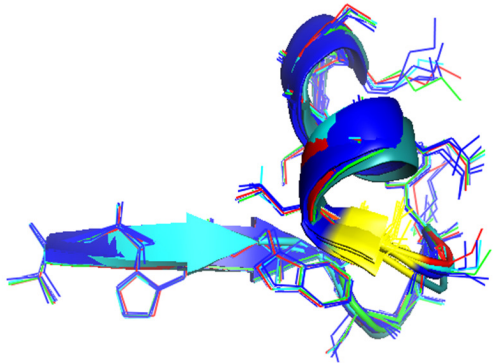

**Figure 1:** **Superposition of ±10 positions around the *S*-site in protein structures.**
In this example, the PDB-IDs are: 1AZM, 1BZM, 1CRM, 1CZM, 1HCB, 1HUG, 1HUH, 1J9W, 1JV0, 2CAB, and 2FOY. The figure was generated using PyMOL software. The Serine residue is colored in yellow.

c. The sequence and taxonomy information for the data used in the analysis are extracted from the UniProtKB database (www.uniprot.org) and shown in columns labeled *Uniprot_Accession_No* and *Protein_Organism* in *glycos_public.xlsx* [38]. The ligand information (see column labeled *Ligand_Sequence* in *glycos_public.xlsx*) is obtained from the PDBsum database [39]

d. Accessible Surface Area (ASA) values (see column labeled *ASA* in *glycos_public.xlsx*) are obtained using the ASAView tool and database [40]. The definitions in the ASAView tool are used to classify the amino acids into five categories: positively charged residues (*R, K, H*), negatively charged residues (*D, E*), polar uncharged residues (*G, N, Y, Q, S, T, W*), Cysteine (*C*), and hydrophobic residues (*L, V, I, A, F, M, P*).

**Ridge Regression Estimated Linear Probability Model Predictions of *O*-glycosylation in Proteins with Structural and Sequence Data**

Table 1: Description of the collected data*

| Dataset name given | Data description | Identified in *glycos_public.xlsx* by | Sample Size | Source |
|---|---|---|---|---|
| *dbogap-str* | *O*-GlcNAc glycosylated sequences with sequence *and* structural data | Oglycos_status = yes | 1,105. Where 998 are human with unique PDB-IDs; of these, only 16 are inferred from known *O*-GlcNAcylated orthologs (the others are experimentally validated). The remaining 107 sequences are non-human. These 998 sequences are in-sample data for the proteins with sequence *and* structural information. The structural information on the 998 sequences was collected | dbOGAP |
| *dbogap* | Human *O*-GlcNAc glycosylated sequences | Ogly_only_seq = yes | 376. These are unique UniProt Accession No. and position pairs. | dbOGAP |
| *dbogap-unique-seq-with-str* | Extract of human sequences from *dbogap-str* with unique UniProt Accession number *and* position pairs (i.e., structure is ignored) | Not identified as it is derivable using software like *SAS* or *R* | 39. Of these, 28 are experimentally validated and the remaining 11 are inferred. These 39 sequences become 998 in *dbogap-str* via the richness in conformational changes associated with them. Of the 39 sequences, 25 are unique proteins (UniProt Accession Nos.) | N/A |
| *dbogap-seq* | Merge *dbogap* with *dbogap-unique-seq-with-str*, by UniProt Accession No. *and* position, and retain those in the first dataset, but not in the second. | Not identified as it is derivable | 340 | N/A |
| *Oglc-PS+* | Additional extract of *O*-GlcNAc glycosylated human sequences | Ogly_21 = yes | 411. Of these, 59.12% are glycosylated at *S* and the others at *T*. Of the 25 unique human proteins in *dbogap-unique-seq-with-str*, 18 are in *Oglc-PS+* | PhosphoSitePlus |
| *Oglc-non-dbogap* | Merge *Oglc-PS+* with *dbogap-seq* and *dbogap-unique-seq-with-str*, by UniProt Accession No. *and* position, and retain those in *Oglc-PS+*, but not in *dbogap-seq* or *dbogap-unique-seq-with-str* | GLCNAC_s1 = yes | 259. This is used as out-of-sample data. Note, 152 of the 340 sequences in *dbogap-seq* are in *Oglc-PS+*. The total number of unique sequences (UniProt Accession No. and UniProt Accession No. position pairs) in *dbogap* and *Oglc-PS+* is 638. | N/A |
| *Ogal* | *O*-GalNAc glycosylated human sequences | GALNAC_s1 = yes | 2,079. This is used as out-of-sample data. Of these, 60.27% are glycosylated at *T* and the others at *S* | PhosphoSitePlus |
| *Ngly* | *N*-glycosylated sequences | glyco_status = yes | 6,328. Of these, 2,422 are "Homo sapiens (Human)". Of the 2,422, the count of sequences with more than one sugar bound is 1,083. These 1,083 sequences are in-sample data for the proteins with sequence *and* structural information. If structure is ignored, there are 361 unique sequences (i.e., unique UniProt Accession No. and position pairs). These 361 sequences are in-sample data for the proteins with only sequence data | Gana et al.[34] |
| *Phosy* | [41]Phosphorylated sequences | Not identified. This is archived in a separate file: *Phosy.csv* | 363,256. Of these, 227,810 are human with amino acids in ±7 positions of the *S/T*-site; and 58.95%, 24.51%, and 16.54% are phosphorylated at *S*, *T* and *Y*, respectively | PhosphoSitePlus |
| *WSTW-Uniprot* | Human sequences with the *W—S/T—W* sequon | wstw = yes | 236. This extract is unique in terms of Uniprot Accession No. & position pairs | UniProt |

\* The columns describe the dataset name, counts of the sequences collected, description of the data and its source. For example, 1,105 *O*-GlcNAc glycosylated proteins with sequence *and* structural data are collected and stored as dataset *dbogap-str*. This data is identified in *glycos_public.xlsx* by "yes" in column *Oglycos_status*. In terms of unique PDB-IDs, there are 998 sequences in this data. The last column cites the source of the collected data, dbOGAP.

e. The ProMotif database is used to obtain the turn types: http://www.ebi.ac.uk/thornton-srv/databases/cgi_in/pdbsum/GetPage.pl?pdbcode=n/a&template=doc_promotif.html. These are in *glycos_public.xlsx* in the column labeled *Turn_Type*.

*Results from examining the empirical data*

The distribution of groups of residues around the *O*-GlcNAc glycosylated proteins, in dataset *dbogap-seq*, is shown in Table 2. It is striking that the immediate vicinity of the *S/T*-site is predominantly polar uncharged and hydrophobic at position −1 and +1, respectively.

The marginal distributions of residues in the vicinity of the *O*-GlcNAc and *O*-GalNAc glycosylated sequences are shown in Table 3 and Table 4, respectively. About 10% or more of the sequences in positions −4 thru −8 have *S*, *P*, or *A* in Table 3; positions −2 and −3 are proline rich; and amino acids *C*, *F*, *M*, and *W* do not occur much. There are some differences between amino acid preferences in *O*-GlcNAc and *O*-GalNAc glycosylated sequences. As reported by earlier studies, the vicinity of the *S/T*-site in *O*-GalNAc glycosylated sequences is proline rich.

The marginal distributions of residues in phosphorylated sequences are shown in Table 5. Along the sequence, *S* tends to show up the most in phosphorylated sequences. There are higher percentages of charged amino acids in these sequences, as well.

**Table 2**: **Percentage distribution of amino acids grouped according to their physiochemical properties around the O-GlcNAc glycosylated S/T-site in the data**

| Amino Acid Position | Positively charged | Negatively charged | Polar uncharged | Cystein | Hydrophobic |
|---|---|---|---|---|---|
| −8 | 14.71 | 8.53 | 40.88 | 0.59 | 35.29 |
| −7 | 13.82 | 8.82 | 36.47 | 1.47 | 39.41 |
| −6 | 14.12 | 7.94 | 42.35 | 0.59 | 35.00 |
| −5 | 18.53 | 8.82 | 32.35 | 0.59 | 39.71 |
| −4 | 18.82 | 8.24 | 35.59 | 0 | 37.35 |
| −3 | 10.88 | 5.00 | 30.00 | 0.88 | 53.24 |
| −2 | 14.12 | 4.41 | 28.82 | 0.29 | 52.35 |
| −1 | 8.24 | 3.82 | 38.24 | 0.59 | 49.12 |
| 0 (*S/T*-site) | | | | | |
| +1 | 10.59 | 8.24 | 52.35 | 0.29 | 28.53 |
| +2 | 6.47 | 6.47 | 45.29 | 0.29 | 41.47 |
| +3 | 8.82 | 4.71 | 44.71 | 0.59 | 41.18 |
| +4 | 10.88 | 6.18 | 48.24 | 0 | 34.71 |
| +5 | 16.47 | 5.00 | 32.35 | 0.59 | 45.59 |
| +6 | 16.47 | 6.18 | 37.06 | 0.29 | 40.00 |
| +7 | 15.88 | 6.76 | 44.41 | 0.29 | 32.65 |
| +8 | 14.41 | 6.47 | 42.94 | 0.59 | 35.59 |

**Table 3:** Marginal distributions (%) of amino acids by position relative to the *S/T*-site

| Amino Acid | Eight positions to the left/right of the *S/T*-site for O-GlcNAc glycosylated sequences | | | | | | | | | | | | | | | |
|---|---|---|---|---|---|---|---|---|---|---|---|---|---|---|---|---|
| | -8 | -7 | -6 | -5 | -4 | -3 | -2 | -1 | +1 | +2 | +3 | +4 | +5 | +6 | +7 | +8 |
| **A** | 9.1 | 12.7 | 7.1 | 12.1 | 11.5 | 8.8 | 11.8 | 10.3 | 9.1 | 17.4 | 12.1 | 8.5 | 10.3 | 10.6 | 7.4 | 7.1 |
| **C** | 0.6 | 1.5 | 0.6 | 0.6 | 0 | 0.9 | 0.3 | 0.6 | 0.3 | 0.3 | 0.6 | 0 | 0.6 | 0.3 | 0.3 | 0.6 |
| **D** | 4.4 | 2.7 | 2.7 | 3.8 | 4.1 | 1.5 | 2.4 | 2.7 | 2.9 | 3.5 | 2.1 | 2.4 | 3.2 | 2.4 | 3.8 | 2.4 |
| **E** | 4.1 | 6.2 | 5.3 | 5.0 | 4.1 | 3.5 | 2.1 | 1.2 | 5.3 | 2.9 | 2.7 | 3.8 | 1.8 | 3.8 | 2.9 | 4.1 |
| **F** | 1.5 | 1.8 | 2.9 | 1.5 | 2.4 | 2.4 | 1.8 | 1.8 | 2.4 | 1.2 | 1.8 | 1.5 | 2.7 | 2.4 | 1.2 | 1.8 |
| **G** | 8.5 | 6.2 | 5.9 | 5.9 | 5.3 | 6.8 | 6.8 | 5.6 | 9.4 | 8.5 | 7.1 | 9.1 | 5.6 | 6.5 | 8.8 | 5.9 |
| **H** | 3.2 | 0.9 | 2.1 | 0.6 | 3.5 | 2.7 | 1.8 | 1.5 | 1.5 | 2.7 | 1.5 | 1.2 | 2.1 | 3.2 | 1.5 | 3.2 |
| **I** | 4.4 | 4.7 | 3.2 | 4.1 | 3.5 | 4.4 | 2.9 | 4.4 | 2.9 | 2.1 | 3.5 | 3.2 | 5.9 | 3.8 | 1.8 | 2.9 |
| **K** | 7.1 | 6.5 | 5.3 | 8.2 | 7.1 | 3.8 | 7.1 | 3.8 | 4.4 | 1.2 | 2.4 | 5.0 | 7.4 | 5.9 | 7.4 | 5.9 |
| **L** | 7.4 | 5.6 | 6.5 | 10.0 | 5.9 | 5.6 | 7.1 | 5.0 | 5.6 | 4.1 | 7.7 | 5.3 | 7.7 | 6.2 | 5.9 | 6.8 |
| **M** | 1.2 | 1.5 | 1.8 | 1.5 | 2.1 | 0.3 | 1.2 | 0.9 | 0.6 | 1.2 | 0.3 | 1.8 | 2.4 | 0.9 | 1.5 | 1.2 |
| **N** | 2.9 | 1.8 | 3.5 | 0.9 | 1.5 | 2.1 | 1.8 | 0.3 | 1.5 | 2.1 | 3.8 | 1.8 | 1.2 | 2.1 | 1.8 | 2.4 |
| **P** | 6.2 | 8.2 | 8.5 | 6.5 | 7.7 | 17.9 | 19.4 | 6.8 | 2.7 | 8.8 | 9.1 | 6.5 | 7.9 | 11.8 | 7.7 | 8.8 |
| **Q** | 4.7 | 2.4 | 5.6 | 3.2 | 4.7 | 2.7 | 5.6 | 2.7 | 9.1 | 5.6 | 7.9 | 2.7 | 3.8 | 5.6 | 6.2 | 6.2 |
| **R** | 4.4 | 6.5 | 6.8 | 9.7 | 8.2 | 4.4 | 5.3 | 2.9 | 4.7 | 2.7 | 5.0 | 4.7 | 7.1 | 7.4 | 7.1 | 5.3 |
| **S** | 14.4 | 13.2 | 14.7 | 10.0 | 12.9 | 6.8 | 8.8 | 12.1 | 16.5 | 19.4 | 14.7 | 19.1 | 8.5 | 11.8 | 13.2 | 15.0 |
| **T** | 6.8 | 8.8 | 9.7 | 8.8 | 8.2 | 8.5 | 4.7 | 13.8 | 13.8 | 8.8 | 8.8 | 13.5 | 10.0 | 7.1 | 11.8 | 10.0 |
| **V** | 5.6 | 5.0 | 5.0 | 4.1 | 4.4 | 13.8 | 8.2 | 20.0 | 5.3 | 6.8 | 6.8 | 7.9 | 8.8 | 4.4 | 7.4 | 7.1 |
| **W** | 0.6 | 0.9 | 1.2 | 0.3 | 0.3 | 0.3 | 0 | 0 | 0 | 0 | 0.3 | 0 | 0 | 0 | 0.9 | 0.6 |
| **Y** | 2.9 | 3.2 | 1.8 | 3.2 | 2.7 | 2.9 | 1.2 | 3.8 | 2.1 | 0.9 | 2.1 | 2.1 | 3.2 | 4.1 | 1.8 | 2.9 |

**Table 4:** Marginal distributions (%) of amino acids by position relative to the *S/T*-site

| Amino Acid | Eight positions to the left/right of the *S/T*-site for O-GalNAc glycosylated sequences | | | | | | | | | | | | | | | |
|---|---|---|---|---|---|---|---|---|---|---|---|---|---|---|---|---|
| | -8 | -7 | -6 | -5 | -4 | -3 | -2 | -1 | +1 | +2 | +3 | +4 | +5 | +6 | +7 | +8 |
| **A** | 7.7 | 7.2 | 7.0 | 7.2 | 6.6 | 7.0 | 9.1 | 8.5 | 7.7 | 7.7 | 10.2 | 7.9 | 6.8 | 8.0 | 8.2 | 7.9 |
| **C** | 0.9 | 0.8 | 0.9 | 0.4 | 0.8 | 0.4 | 0.4 | 0.1 | 0.9 | 0.1 | 0.3 | 0.4 | 0.7 | 0.6 | 0.7 | 1.4 |
| **D** | 4.3 | 4.5 | 4.1 | 3.8 | 4.1 | 3.4 | 4.2 | 1.3 | 3.9 | 4.0 | 4.3 | 3.7 | 5.2 | 4.6 | 4.6 | 5.0 |
| **E** | 6.4 | 6.2 | 6.6 | 6.8 | 6.4 | 8.3 | 6.6 | 4.4 | 7.0 | 6.2 | 5.2 | 7.0 | 7.8 | 6.6 | 7.1 | 8.0 |
| **F** | 2.4 | 2.4 | 2.3 | 2.2 | 2.3 | 2.4 | 2.7 | 2.8 | 1.6 | 2.0 | 2.7 | 2.0 | 2.3 | 2.3 | 2.0 | 2.2 |
| **G** | 6.6 | 6.7 | 6.9 | 6.8 | 6.2 | 5.2 | 7.6 | 5.3 | 5.6 | 7.2 | 6.0 | 6.8 | 6.1 | 7.6 | 7.8 | 6.8 |
| **H** | 3.6 | 3.8 | 2.9 | 4.3 | 4.5 | 3.6 | 3.6 | 4.5 | 3.8 | 4.0 | 2.8 | 3.5 | 3.5 | 3.2 | 3.4 | 3.0 |
| **I** | 2.9 | 3.2 | 2.7 | 3.0 | 3.6 | 3.4 | 3.0 | 3.2 | 2.4 | 2.7 | 2.3 | 2.7 | 3.2 | 2.1 | 2.0 | 2.3 |
| **K** | 5.7 | 5.4 | 5.7 | 5.8 | 5.1 | 5.6 | 5.5 | 3.3 | 4.6 | 5.6 | 3.6 | 6.0 | 5.3 | 5.6 | 6.6 | 7.0 |
| **L** | 6.8 | 7.1 | 6.7 | 7.2 | 6.8 | 8.0 | 7.1 | 5.6 | 7.5 | 7.1 | 6.1 | 6.7 | 7.2 | 7.7 | 7.1 | 7.0 |
| **M** | 1.7 | 1.3 | 1.0 | 0.8 | 0.8 | 1.0 | 1.1 | 0.9 | 0.7 | 1.0 | 0.9 | 1.0 | 1.1 | 1.0 | 1.1 | 1.1 |
| **N** | 2.1 | 2.0 | 2.5 | 2.4 | 2.1 | 2.0 | 1.2 | 1.7 | 1.7 | 2.6 | 2.1 | 2.0 | 2.2 | 2.2 | 2.3 | 2.6 |
| **P** | 8.3 | 9.6 | 9.4 | 10.2 | 9.4 | 11.0 | 11.5 | 15.8 | 11.6 | 12.2 | 17.8 | 12.3 | 9.7 | 10.6 | 8.9 | 8.2 |
| **Q** | 4.7 | 4.3 | 4.4 | 4.4 | 4.7 | 4.2 | 4.5 | 3.9 | 4.9 | 3.9 | 4.2 | 3.9 | 3.6 | 4.0 | 4.9 | 4.2 |
| **R** | 7.5 | 6.5 | 7.5 | 7.2 | 7.5 | 6.8 | 6.2 | 5.1 | 5.9 | 5.8 | 4.7 | 6.5 | 6.9 | 7.5 | 6.7 | 6.8 |
| **S** | 9.4 | 10.6 | 11.5 | 10.1 | 11.8 | 10.3 | 9.0 | 10.3 | 10.4 | 11.3 | 10.8 | 10.3 | 9.6 | 10.1 | 10.5 | 9.1 |
| **T** | 11.5 | 10.8 | 10.6 | 10.2 | 10.1 | 8.3 | 8.0 | 11.2 | 12.0 | 6.8 | 10.9 | 10.2 | 9.5 | 9.7 | 10.0 | 10.4 |
| **V** | 5.2 | 5.8 | 5.4 | 5.2 | 5.5 | 6.7 | 6.0 | 9.7 | 6.2 | 5.1 | 5.4 | 6.1 | 6.0 | 4.5 | 4.9 | 4.8 |
| **W** | 1.0 | 0.6 | 0.5 | 0.8 | 0.5 | 0.9 | 0.8 | 0.5 | 0.6 | 0.7 | 0.5 | 1.0 | 0.6 | 0.5 | 0.5 | 1.3 |
| **Y** | 1.3 | 1.4 | 1.2 | 1.4 | 1.5 | 1.6 | 1.8 | 1.7 | 1.2 | 1.5 | 1.7 | 1.4 | 1.4 | 1.5 | 1.3 | 1.3 |

**Table 5:** Marginal distributions (%) of amino acids by position relative to the *S/T/Y/H*-site

| Amino Acid | Seven positions* to the left/right of the *S/T*-site for phosphorylated sequences | | | | | | | | | | | | | |
|---|---|---|---|---|---|---|---|---|---|---|---|---|---|---|
| | -7 | -6 | -5 | -4 | -3 | -2 | -1 | +1 | +2 | +3 | +4 | +5 | +6 | +7 |
| *A* | 7 | 6.9 | 6.8 | 6.6 | 6.4 | 6.8 | 6.7 | 5.7 | 6.4 | 6.4 | 6.4 | 6.6 | 6.9 | 6.7 |
| *C* | 1.4 | 1.4 | 1.3 | 1.2 | 1.2 | 1.3 | 1.3 | 1.2 | 1.4 | 1.2 | 1.3 | 1.3 | 1.3 | 2.5 |
| *D* | 4.9 | 4.9 | 4.9 | 5 | 5 | 5.1 | 6.1 | 5.4 | 5.9 | 5.6 | 5.2 | 5.3 | 5.1 | 5.1 |
| *E* | 7.3 | 7.3 | 7 | 7.2 | 6.8 | 6.6 | 6 | 6.6 | 9.5 | 8.5 | 7.8 | 7.6 | 7.8 | 7.5 |
| *F* | 2.6 | 2.6 | 2.8 | 2.6 | 2.4 | 2.5 | 2.8 | 3.1 | 2.3 | 2.5 | 2.6 | 2.7 | 2.7 | 2.7 |
| *G* | 6.9 | 6.7 | 6.6 | 7.1 | 6.8 | 6.7 | 8.1 | 7.6 | 6.9 | 7.1 | 6.5 | 6.6 | 6.6 | 6.5 |
| *H* | 2.2 | 2.2 | 3.3 | 2.2 | 2.5 | 2.1 | 3.7 | 2 | 2 | 2.1 | 2.1 | 2.3 | 2.2 | 2.3 |
| *I* | 3.6 | 3.7 | 3.6 | 3.5 | 3.4 | 4.1 | 3.9 | 3.6 | 3.5 | 3.5 | 3.8 | 3.5 | 3.6 | 3.5 |
| *K* | 7.4 | 7.3 | 7 | 7.3 | 7.2 | 6.1 | 6 | 4.5 | 6 | 7.6 | 6.8 | 7.3 | 7.5 | 7.2 |
| *L* | 7.9 | 8.3 | 8.8 | 7.7 | 8 | 8 | 9.1 | 9.1 | 7.9 | 7.8 | 8.7 | 7.9 | 8 | 8.1 |
| *M* | 2.1 | 1.8 | 1.8 | 1.8 | 1.8 | 1.9 | 1.7 | 1.8 | 1.7 | 1.8 | 1.8 | 1.8 | 1.8 | 1.8 |
| *N* | 3.4 | 3.4 | 3.4 | 3.5 | 3.3 | 3.8 | 3.8 | 2.6 | 3.3 | 3.4 | 3.3 | 3.3 | 3.3 | 3.4 |
| *P* | 7.1 | 7 | 7 | 6.9 | 6.7 | 7.9 | 7.3 | 15.5 | 8 | 7.3 | 8.1 | 7.6 | 7.4 | 7 |
| *Q* | 4.5 | 4.4 | 4.3 | 4.8 | 4.1 | 4.4 | 3.8 | 4.6 | 3.9 | 4.3 | 4.2 | 4.3 | 4.4 | 4.3 |
| *R* | 7.4 | 7.5 | 7.7 | 7.3 | 11 | 7.3 | 6.8 | 5.1 | 6.1 | 6.8 | 6.4 | 6.9 | 7 | 7 |
| *S* | 10.6 | 10.7 | 10.3 | 11.7 | 10.9 | 11.7 | 9.8 | 9.2 | 11.6 | 11.1 | 11.3 | 10.6 | 10.8 | 10.5 |
| *T* | 5.7 | 5.5 | 5.3 | 5.4 | 5 | 5.7 | 4.4 | 4.1 | 5.9 | 5.1 | 5.7 | 5.5 | 5.6 | 5.5 |
| *V* | 5.2 | 5.2 | 5.2 | 5.1 | 4.8 | 5.4 | 5.5 | 5.3 | 5.2 | 5 | 5.3 | 5 | 5.3 | 5.1 |
| *W* | 0.8 | 0.7 | 0.7 | 0.7 | 0.6 | 0.6 | 0.6 | 0.8 | 0.5 | 0.7 | 0.7 | 0.7 | 0.7 | 0.8 |
| *Y* | 2.2 | 2.2 | 2.2 | 2.1 | 2 | 2.1 | 2.3 | 2.1 | 2 | 2.1 | 2.1 | 3.2 | 2.1 | 2.5 |

* Only seven positions to the left/right of the phosphorylation site were collected.

While visual inspection of amino acids in the vicinity of these sites provide some indication of their preferences and charge distributions, a subjective analysis cannot be done. Therefore, statistical analysis of the data is done in this paper to provide more rigorous indications of the same. Empirical analysis suggests that given a set of sequences for a particular PTM of proteins, the proportions of an amino acid occupying positions (e.g., $\pm 8$ from the *S/T*-site) along a sequence can discriminate between two PTMs of proteins. This is tested using the Wilcoxon signed-rank test [41] and the results are shown in Table 6.

**Table 6:** Testing whether the proportions of an amino acid over the positions it occupies are pairwise different across PTMs of proteins*

| | Amino Acid | | | | | | | | | | | | | | | | | | | |
|---|---|---|---|---|---|---|---|---|---|---|---|---|---|---|---|---|---|---|---|---|
| | *A* | *C* | *D* | *E* | *F* | *G* | *H* | *I* | *K* | *L* | *M* | *N* | *P* | *Q* | *R* | *S* | *T* | *V* | *W* | *Y* |
| *O-GlcNAc* vs *O-GalNAc* glycosylation | 1 | 0 | 1 | 1 | 1 | 0 | 1 | 1 | 0 | 0 | 0 | 0 | 1 | 0 | 0 | 1 | 0 | 1 | 1 | 1 |
| *O-GlcNAc* glycosylation vs phosphorylation | 1 | 1 | 1 | 1 | 1 | 0 | 0 | 0 | 1 | 1 | 1 | 1 | 0 | 0 | 0 | 0 | 1 | 1 | 1 | 0 |
| *O-GalNAc* glycosylation vs phosphorylation | 1 | 1 | 1 | 1 | 1 | 0 | 1 | 1 | 1 | 1 | 1 | 1 | 1 | 0 | 0 | 0 | 1 | 1 | 0 | 1 |

* 1/0 indicates the difference is/isn't statistically significant at the 5% level of significance. The first comparison uses the 16 pairs, while the other two uses 14 pairs.

For example, in Table 3 and Table 5, the proportions of occurrence of amino acid *F* along seven positions to the left/right of the *S/T*- and *S/T/Y/H*- sites in *O*-GlcNAc glycosylated and phosphorylated sequences, respectively, appear different under pairwise comparison (i.e., the amino acid proportions along one row of one Table are compared, pairwise, with the proportions of that same amino acid along the corresponding row of the other Table). The Wilcoxon signed-rank test is used to determine if these two sets of proportions are statistically different. As shown in Table 6, detected differences are flagged as "1" and by "0" otherwise. Table 6 indicates that using amino acids in the positions they occupy as "explanatory variables" to predict *O*-glycosylation is reasonable. In contrast, the proportions of amino acids in a particular position for one PTM process are not statistically different from the proportions of amino acids in that same position for another PTM process. For example, the Wilcoxon signed-rank test applied to compare, pairwise, the proportions in any column of Table 3 with those in the corresponding column of Table 4 shows the pairwise differences are not statistically significant. This same result holds when comparing the columns of Table 3 with the corresponding ones in Table 5; and those of Table 4 with the corresponding ones in Table 5.

Finally, it is of interest to know whether sequons involved in the PTMs of proteins are unique to each kind of PTM process. For example, given the consensus sequon for *N*-glycosylation is *N–X–S/T*, one can find out if in the collected data there are *N*-glycosylated sequences that are also *O*-glycosylated or phosphorylated at the *S/T*-site, which is at the 2[nd] position to the right of *N* in this sequon. UniProt Accession numbers and UniProt position pairs are compared across PTMs. Because *N*-glycosylation occurs at *N*, the UniProt positions of *N*-glycosylated sequences, in the data, are increased by 2 so the *S/T*-sites can be compared across the other PTMs. In the empirical data, there are no *N*-glycosylated sequences with UniProt Accession No. and UniProt position + 2 pairs that match the UniProt Accession No. and UniProt position pairs of *O*-GlcNAc or *O*-GalNAc glycosylated sequences. This is as expected, because the structures of *N*- and *O*-linked oligosaccharides are very different, and reflect differences in their biosynthesis and cellular localizations. Thus, the *N*-linked site cannot be *O*-glycosylated at the *S/T*-site of the *N–X–S/T* sequon. However, there is a small number of *N*-glycosylated sequences that are phosphorylated. Table 7 presents the counts of sequences in the empirical data that are common between the different PTMs of proteins considered herein (i.e., intersections of two sets of sequences are calculated).

**Table 7:** UniProt Accession No. *and* position pair counts that are common between PTMs of proteins

| Sequences in the empirical data that are: | *N*-glycosylated | *O*-GlcNAc glycosylated | *O*-GalNAc glycosylated | Phosphorylated |
|---|---|---|---|---|
| *N*-glycosylated | 361 | 0 | 0 | 13 |
| *O*-GlcNAc glycosylated | | 638 | 0 | 144 |
| *O*-GalNAc glycosylated | | | 2,079 | 224 |
| Phosphorylated | | | | 227,810 |

Table 8 lists 13 proteins that are *N*-glycosylated at *N* and phosphorylated at the second position to the right of *N*, which is *S/T*. The majority of these proteins are receptors involved in signaling and are phosphorylated via a threonine residue. What roles these phosphorylation sites play require further analysis and experimentation.

**Table 8**: *N*-glycosylated sequences that are also phosphorylated

| UniProt Accession No. | Protein Name | Phosphorylated at UniProt position (residue) | *N*-glycosylated at UniProt position |
|---|---|---|---|
| O00206 | Toll-like receptor 4 | 175 (*T*) | 173 |
| O15455 | Toll-like receptor 3 | 72 (*T*) | 70 |
| P00533 | Epidermal growth factor receptor | 354 (*T*) | 352 |
| P02675 | Fibrinogen beta chain | 396 (*T*) | 394 |
| P02788 | Lactotransferrin | 499 (*T*) | 497 |
| P04629 | High affinity nerve growth factor receptor | 123 (*S*) | 121 |
| P05187 | Alkaline phosphatase | 273 (*T*) | 271 |
| P06213 | Insulin receptor | 366 (*S*) | 364 |
| P06213 | Insulin receptor | 447 (*T*) | 445 |
| P07711 | Cathepsin L1 | 223 (*T*) | 221 |
| P12821 | Angiotensin-converting enzyme | 716 (*T*) | 714 |
| Q96FE5 | Leucine-rich repeat and immunoglobulin-like domain-containing Nogo receptor-interacting protein 1 | 295 (*S*) | 293 |
| Q9NPH3 | Interleukin-1 receptor accessory protein | 113 (*T*) | 111 |

## *Modeling Strategy*

Because structural data on *O*-glycosylated sequences is sparse, two LPMs are estimated. One with sequence *and* structural data as explanatory variables. The other with only sequence data. The LPM has two principal sources of error: the exact process by which a protein becomes *O*-glycosylated is unknown; and uncertainty due to ambiguity in what constitutes a set of proteins that cannot be *O*-glycosylated. So, a suitable sequon is needed for discriminating between the two outcomes.

Data (see Table 1) on three PTMs of proteins, increasing in sample size, is collected to identify a consensus sequon: *O*-GlcNAc glycosylation, *O*-GalNAc glycosylation, and phosphorylation, which is an important and pervasive PTM of proteins [42] that occurs, generally, at the *S/T/Y*-site. It has also been found to occur at the *H*-site [43]. Several sequons are tested to see if a consensus sequon exists. The outcome of these tests is shown in Table 9. As can be noted, the sequon *W – S/T – W* does not occur in the collected set of *O*-glycosylated sequences. Thus, ~ (*W – S/T – W*) is likely a consensus composite sequon for *O*-glycosylation, where, following *Principia Mathematica* [44], "~" denotes the "not" operator. This means there are 3 sequons that satisfy ~ (*W – S/T – W*): ~*W – S/T – W*,   *W – S/T – ~W*,   or   ~*W – S/T – ~W*.

An important question is whether, as sample size increases, *W – S/T – W* continues not to occur *at all* in *O*-glycosylated sequences. The answer to this question has to wait until much larger sets of *O*-glycosylated sequences are available for study. In the meantime, one can approach this question in another way. The *N – ~P – S/T* sequon is considered a necessary [45, 46], but not sufficient, condition for *N*-glycosylation. This is another important PTM of proteins. Its likelihood is modeled in Gana et al. In the strictest sense, this is a true statement if the occurrence of *P* in the position between *N* and *S/T* is an "absolute" inhibitor of *N*-glycosylation [47] – i.e., if the *N – ~P – S/T* sequon is never observed in *N*-glycosylated sequences. However, this is not the case. Rather, the word "absolute" is used in a probabilistic sense: the fact that *N – P – S/T* occurs "rarely" is sufficient for *N – ~P – S/T* to be termed a "consensus" sequon. In particular, a partial search for the *N–P–S/T* sequons in UniProt revealed 23 such sequons associated with *N*-glycosylated human proteins. In contrast, 16 (0.0066% + 0.004% = 0.007%) of the 227,810 phosphorylated human sequences in Table 9 have the $W – S/T/Y/H – W$ sequon. Finally, a search of UniProt for human proteins with the $W – S/T – W$ sequon yielded 236 sequences (in the *WSTW-Uniprot* dataset of Table 1). It is found that none of them are *O*-glycosylated at *S/T*.

The sequon for *C*-linked glycosylation is $W–X–X–W$ or $W–S/T–X–C$. Analysis of the data indicates that the presence of $W$ immediately before and immediately after the *S/T*-site (*S/T/Y/H*-site) disallows *O*-glycosylation (phosphorylation). Furthermore, the occurrence of $N$ two positions to the left of the *S/T*-site is relatively rare in *O*-glycosylation. Thus, nature appears to have evolved distinct sequons and amino acid propensities unique to each of the PTMs discussed herein.

Table 9 indicates that the majority (98.54% + 0.24% = 98.78%) of *O*-GlcNAc glycosylated sequences have the $\sim N – X – S/T$ sequon. In contrast, these sequences having the $N – \sim P – S/T$ sequon is small (1.22%). Thus, for LPM estimation, the set of *N*-glycosylated sequences in Gana et al. [34] is considered to be a reasonable approximation for non-*O*-glycosylated sequences. By discriminating between the features of *O*- and non-*O*-glycosylated sequences, respectively, the LPM estimates the likelihood of *O*-glycosylation. Because 1.22%, rather than 0%, of the *O*-glycosylated sequences have the $N – \sim P – S/T$ sequon, the ambiguity regarding what constitutes a set of sequences that cannot be *O*-glycosylated has not been completely removed. However, that ambiguity has been considerably reduced.

**Table 9: Empirical occurrence rate of the identified sequon in PTMs of proteins**

| Sequon* (viewed from *S/T/Y/H* as "center") | % of Sequences in the collected data that are: | | |
|---|---|---|---|
| | *O*-GlcNAc glycosylated | *O*-GalNAc glycosylated | Phosphorylated |
| $N – P – S/T$ | 0.24% | 0.43% | 0.145% |
| $N – \sim P – S/T$ | 1.22% (0.5%, 2.8%) ** | 0.77% (0.5%, 1.2%) ** | 2.858% (2.8%, 2.9%) ** |
| $\sim N – X – S/T$ | 98.54% (96.9%, 99.3%) ** | 98.80% (98.2%, 99.2%) ** | 80.453% |
| $\sim N – X – Y$ | Not applicable | Not applicable | 15.748% |
| $\sim N – X – H$ | Not applicable | Not applicable | 0.006% |
| $\sim N – X – S/T/Y/H$ | Not applicable | Not applicable | 96.2% (96.1%, 96.3%) ** |
| $W – S/T – W$ | 0% | 0% | 0.0066% (0.004%, 0.011%) ** |
| $\sim W – S/T – \sim W$ | 100% | 98.94% (98.4%, 99.3%) ** | 82.3133% |
| $\sim W – S/T – W$ | 0% | 0.58% (0.33%, 1.0%) ** | 0.6422% |
| $W – S/T – \sim W$ | 0% | 0.48% (0.26%, 0.88% ** | 0.4930% |
| $W – Y/H – W$ | Not applicable | Not applicable | 0.0004% |
| $\sim W – Y/H – \sim W$ | Not applicable | Not applicable | 16.2587% |
| $W – Y/H – \sim W$ | Not applicable | Not applicable | 0.1365% |
| $\sim W – Y/H – W$ | Not applicable | Not applicable | 0.1462% |
| | | | |
| *Sequence count* | *411* | *2,079* | *227,810* |

*X denotes any amino acid. ** 95% confidence interval [48]

Using the data in Gana et al., *O*-GlcNAc glycosylation is modeled by including the amino acids in 8 positions to the right, and 8 positions to the left, of the *S/T*-site as possible explanatory variables. The collected data on *N*-glycosylated sequences described in Gana et al. [34] have 10 positions to the right, and 10 positions to the left, of the *N*-site. Because training the LPM to discriminate between *O*- and not *O*- glycosylated proteins must focus on the *S/T*-site as center for all proteins in the estimation dataset, only 8 positions to the right of *S/T* are available for use in that data. The remaining possible explanatory variables are the collected structural data on the proteins.

When the LPM is applied to out-of-sample data, other types of *O*-glycosylated sequences may be a part of it. To assess the sensitivity of the mispredictions rate, the LPM of *O*-GlcNAc glycosylation is applied to a sample of *O*-GalNAc glycosylated sequences. The modeling and validation strategy is summarized in Table 10.

**Table 10: Summary of the modeling/validation strategy**

| Row | Data used | What the data is used for | Outcome |
|---|---|---|---|
| 1 | The 998 sequences in *dbogap-str* and the 1,083 sequences in *glycos* | Predicting the likelihood of *O*-GlcNAc glycosylation with sequence *and* structural data | |
| 2 | The 340 sequences in *dbogap-seq* and the 361 sequences in *Ngly* | Predicting the likelihood of *O*-GlcNAc glycosylation with *only* sequence data. A sequence is considered to be mispredicted if its predicted probability of *O*-glycosylation is less than 50% and it is *O*-glycosylated | Table 14. About 11% of sequences in *dbogap-seq* are mispredicted as not being *O*-GlcNAc glycosylated; and 9% of the sequences in *Ngly* are mispredicted as being *O*-GlcNAc glycosylated |
| 3 | The 259 sequences in *Oglc-non-dbogap* | Calculating the out-of-sample mispredictions rate with the LPM estimated for the exercise outlined in Row 2 of this Table | 54 of the 259 sequences ($\approx$ 21%) are mispredicted as not being *O*-GlcNAC glycosylated |
| 4 | The 2,079 sequences in *Ogal* | Calculate the out-of-sample mispredictions rate with the LPM estimated for the exercise outlined in Row 2 of this Table | 656 of the 2,079 ($\approx$ 31.6%) are mispredicted as not being *O*-GalNAc glycosylated |
| 5 | The 236 sequences in *WSTW-Uniprot* | To see if any of these are *O*-glycosylated | None are *O*-glycosylated. This again indicates that $\sim (W - S/T - W)$ is likely necessary for *O*-glycosylation |

## Estimating the two LPMs

### *LPM predictions using sequence and structural data*

Based on the collected data (*glycos_public.xlsx*), as in Gana et al.[34], the $i^{\text{th}}$ position to the left (right) of *S/T* is *minus$_i$* (*plus$_i$*). If amino acid $\alpha$ occupies $i$, then: *minus$_{i\alpha}$* = 1 if $\alpha$ is to the left of *S/T* and *plus$_{i\alpha}$* = 1 if $\alpha$ is to the right of *S/T*; otherwise, these indicator variables take the value 0. Thus, 140 (20 × 8 − 20) and 160 (20 × 8) indicator variables define the left and right neighborhoods of *S/T*, respectively. The ratio (*pos$_j$*) of the position of the glycosylation site (*S/T* or *N*) to the total length of protein *j* is also assumed to be an explanatory variable. Finally, it is assumed that the structural information on the sequences are additional explanatory variables.

Let $Y_j$ be the binary dependent variable that takes the value 1 if sequence $j$ is *O*-glycosylated and 0 if it is not (*N*-glycosylated). Then, the postulated LPM (indexed as *EQ1*) is:

$Y_j = \beta_0 + \Sigma_\alpha \Sigma_{i \neq 2} \beta_{i\alpha} \, minus_{i\alpha} + \Sigma_\alpha \Sigma_i \varphi_{i\alpha} \, plus_{i\alpha} + \lambda \, pos_j + \delta \, ASA_j + \pi \, (ASA\_zero_j) +$

$\Sigma_k \theta_k \, (Turn \, Type_k) + \omega_1 \, (psi \, angle) + \omega_2 \, (phi \, angle) \rightarrow \rightarrow \rightarrow \rightarrow \rightarrow (EQ1)$

where $\beta_0$, $\beta_{i\alpha}$, $\varphi_{i\alpha}$, $\lambda$, $\delta$, $\pi$, $\theta_k$, $\omega_1$ and $\omega_2$ are constants (coefficients) estimated using the data; $\Sigma_\alpha$, $\Sigma_i$ denote summations over significant amino acid and position combinations, respectively, and $\Sigma_k$ denotes summation over significant turn types. The variable *ASA_zero* is a dummy variable that takes the value 1 if ASA is zero, and 0 otherwise. *Minus$_{2\alpha}$* is dropped so as not to bias the LPM, because the set of non-*O*-glycosylated proteins only has an *N* (i.e., there is no amino acid variation, by design) at minus 2. Thus, retaining *Minus$_{2\alpha}$* with $\alpha \neq N$ will flag amino acids (other than *N*) present at minus 2 only whenever $Y_j = 1$, which would be inappropriate. Amino acids *A*, *D*, *F* and *K* dominate the 998 human proteins at minus 2; and *D*, *F*, *K*, *R* and *S* dominate the remaining 107 non-human proteins. The set of *O*-glycosylated sequences have no *N* at minus 2. Also, note that *OOS_wstw* is not used as the set of non-*O*-glycosylated sequences, because it will lead to the perfect, but trivial, model: $Y_j = 1 - minus_{1W}$. In this case, one option is to estimate the LPM by ignoring the amino acids on either side of the *S/T*-site. However, this does not improve LPM predictive capability materially. More importantly, it would be of interest to find sequences with the *W–S/T–W* sequon that do "something" (e.g., *C*-glycosylation with *W–S/T–W–W* or *W–S/T–W–C* sequons) rather than "nothing" (or staying "silent"); and then refit the LPM to see how well it discriminates between *O*-glycosylated and non-*O*-glycosylatable sequences. In addition, if structural data is included for LPM estimation, solid predictive power would likely be the outcome.

The selected LPM, shown in , is estimated on 1,083 (human) *N*-glycosylated proteins and 987 *O*-glycosylated proteins. As in Gana et al. [34], *N*-glycosylated sites with a single sugar are deleted from the estimation data because these may be false positives given the known artifacts from crystallization. As detailed in Gana et al. [34], the LPM is selected using the stepwise selection method [49-51] in conjunction with 10-fold cross-validation [52-56] to minimize the prediction residual sum of squared errors (PRESS), which has a point of contact with the Brier Score [57, 58] . Because the LS estimated LPM produces heteroscedastic errors (i.e., residuals with varying variances), it was confirmed that stepwise selected significant variables continued to remain significant when reevaluated in terms of heteroscedasticity-consistent standard errors [59], as a first approximation. Because the exact functional form of the LPM error variance is known, RR is applied to the stepwise selected LPM to reconfirm that all variables continue to be statistically significant. This is done by re-estimating the stepwise selected LPM by RR so the predicted values of $Y_j$ are forced to lie in [60] the 0-1 interval and estimating the error variances [29]: (RR predicted $Y_j$ ) $\times$ (1 – RR predicted $Y_j$ ). The optimal value of the RR tuning parameter, $k$, for which all of the aforesaid predictions fall in the range 0-1 is 3.77895. Using these variances as weights, the LPM is re-estimated using classical weighted LS (WLS). If variables with *p*-values greater than 5% result, the entire LPM estimation process is redone until all variables are significant at 5%.

The number of amino acid and position combinations found to be significant in the selected

LPM is 76. Model parsimony (e.g., favoring retaining fewer, rather than more, explanatory variables) is not sought, because it is not at all clear that Nature is "inefficient", or provides redundant information, by making irrelevant the placement of certain amino acids in certain positions along the sequence. The selected LPM yields an in-sample Kolmogorov-Smirnov (KS) statistic of about 99.2%, which signals a reasonable degree of discrimination between *O*- and non-*O*- glycosylated proteins. The maximum separation for the KS statistic [61-65] occurs at the predicted probability value of about 43%. The average (in-sample) predicted probabilities of *O*-glycosylation when $Y_j = 0$ and $Y_j = 1$ are about 1% and 98%, respectively. The standard deviations of predicted probabilities of *O*-glycosylation when $Y_j = 0$ and $Y_j = 1$ are about 11% and 7%, respectively. The Durbin-Watson (DW) statistic [66, 67] for the LPM is about 1.5.

**Table 11: LPM predicting *O*-glycosylation probabilities given structural and sequence data**

| Variable* | β | |t| | Variable* | β | |t| | Variable* | β | |t| |
|---|---|---|---|---|---|---|---|---|
| intercept | -0.1309 | 9.71 | m8N | -0.0921 | 4.18 | p7T | -0.1431 | 9.24 |
| m1D | 0.1878 | 10.74 | m8P | -0.1516 | 11.05 | p7V | -0.1309 | 8.14 |
| m1L | 0.0861 | 7.26 | m8V | -0.2180 | 12.08 | p7Y | 0.1290 | 6.75 |
| m1P | 0.7885 | 11.63 | p1A | 0.1778 | 9.83 | p8H | 0.1164 | 5.76 |
| m1R | -0.0757 | 4.66 | p1D | 0.0915 | 4.67 | p8K | 0.0570 | 3.34 |
| m3A | 0.2162 | 14.35 | p1F | 0.0858 | 5.44 | p8N | 0.1003 | 5.51 |
| m3C | -0.1315 | 6.45 | p1S | 0.0835 | 5.33 | p8Q | 0.1493 | 8.53 |
| m3L | -0.1139 | 9.49 | p1T | 0.1673 | 10.15 | pos | 0.2821 | 15.95 |
| m3N | -0.2245 | 5.86 | p1V | 0.0669 | 4.97 | ASA_zero | 0.8057 | 12.02 |
| m3T | -0.0881 | 5.14 | p2A | 0.0590 | 2.92 | II | 0.1649 | 6.95 |
| m4E | 0.1296 | 5.3 | p2H | 0.3143 | 11.32 | II´ | -0.2828 | 7.58 |
| m4F | -0.0848 | 5.56 | p2P | 0.2652 | 18.99 | Helix | 0.1610 | 14.88 |
| m4N | 0.1419 | 7.3 | p2Q | -0.1532 | 7.89 | Beta Bridges | 0.6077 | 6.53 |
| m4R | 0.1410 | 7.04 | p2Y | -0.2434 | 8.12 | Beta Hairpin | 0.7884 | 22.3 |
| m4V | -0.0545 | 4.18 | p3N | -0.1615 | 7.69 | Beta Hairpin Strand | -0.1044 | 9.83 |
| m5D | 0.2521 | 13.74 | p3W | -0.1086 | 3.41 | Phi angle | -0.0004 | 5.31 |
| m5F | -0.1257 | 6 | p4A | 0.1629 | 9.62 | | | |
| m5G | 0.0700 | 3.21 | p4E | 0.1757 | 12.53 | | | |
| m5I | -0.0981 | 4.85 | p4P | 0.1921 | 13.35 | | | |
| m5Y | -0.1468 | 6.9 | p5C | -0.1957 | 9.63 | | | |
| m6E | 0.0954 | 7.08 | p5E | 0.1718 | 7.98 | | | |
| m6H | -0.1245 | 4.73 | p5H | 0.1938 | 7.88 | | | |
| m6V | 0.1528 | 9.82 | p5I | -0.0760 | 4.33 | | | |
| m6W | 0.1166 | 4.9 | p5Q | 0.1060 | 4.93 | | | |
| m6Y | -0.1265 | 6.31 | p5T | 0.0722 | 4.61 | | | |
| m7A | 0.1576 | 10.18 | p5Y | -0.1876 | 7.45 | | | |
| m7E | -0.1132 | 6.87 | p6F | -0.1155 | 5.43 | | | |
| m7G | -0.0685 | 4.89 | p6G | -0.1696 | 10.07 | | | |
| m7H | 0.2427 | 11.33 | p6M | 0.3118 | 9.57 | | | |
| m7I | 0.1222 | 7.67 | p6N | -0.1209 | 6.67 | | | |
| m7K | 0.1538 | 8.77 | p6Q | -0.2641 | 12.26 | | | |
| m7S | 0.0826 | 4.44 | p7A | 0.1752 | 7.04 | | | |
| m8G | 0.1032 | 6.68 | p7C | -0.1281 | 5.86 | | | |
| m8L | -0.1541 | 10.43 | p7E | -0.1161 | 7.36 | | | |
| m8M | -0.1234 | 5.34 | p7G | -0.0951 | 5.81 | | | |

\* *mi*α and *pi*α are abbreviations for *minus*$_{i\alpha}$ and *plus*$_{i\alpha}$, respectively. Note that the coefficient standard error is β ÷|t|.

The nine mispredictions made by the LPM, in-sample, are documented in Table 12. The accessions in Table 12 are not annotated in UniProt to be glycosylated except for Q16566 (*O*-linked), Q92854 (*N*-linked), and P31749 (phosphorylated). Thus, how many sequences in Table 12 are truly false positives, remains to be seen.

**Table 12: LPM mispredictions in-sample, using 50% as the cutoff probability**

| UniProt Accession No. | UniProt Position | Protein Length | Sequence | Secondary Structure | Turn Type | Phi angle | Psi angle | ASA | LPM prediction |
|---|---|---|---|---|---|---|---|---|---|
| *Proteins with LPM predictions < 50%, but that are deemed to be O-glycosylated in the data* | | | | | | | | | |
| P02730 | 224 | 911 | ILEKIPPD**S**EATLVLVG | Beta Turn | IV | -113 | 168 | 18.8 | 19.3% |
| Q16566 | 57 | 473 | ESELGRGA**T**SIVYRCKQ | Beta Turn | I | -83 | -35 | 69.9 | 34.6% |
| P16157 | 794 | 1881 | LKVVTDET**S**FVLVSDKH | Gamma Turn, Beta Turn | Inverse, IV | -54 | 11 | 94.7 | 36.3% |
| P68431 | 11 | 136 | RTKQTARK**S**TGGKAPRK | Loop | N/A | -81 | 145 | 44.4 | 36.3% |
| P31749 | 308 | 480 | IKDGATMK**T**FCGTPEYL | Loop | N/A | -115 | -3 | 47.6 | 44.5% |
| P04406 | 229 | 335 | IPELNGKL**T**GMAFRVPT | Strand | N/A | -157 | 171 | 55.5 | 44.5% |
| *Proteins with LPM predictions > 50%, but that are deemed to be N-glycosylated in the data* | | | | | | | | | |
| P06756 | 488 SA | 1048 | SILNQDNK**T**CSLPGTAL | Gamma Turn, Beta Turn | Inverse, IV | -69 | 93 | 63.5 | 51.6% |
| P04629 | 329 | 796 | THVNNGNY**T**LLAANPFG | Beta Hairpin Strand | N/A | -93 | 119 | 30.7 | 62.7% |
| Q92854 | 329 | 862 | SAVCAYNL**S**TAEEVFSH | Beta Hairpin Strand | N/A | -76 | 124 | 35.5 | 70.1% |

It has been observed that amino acid *P* is usually present around *O*-glycosylation sites; and that *O*-glycosylation tends to occur more in the beta-strands of proteins [68, 69] indicating the significance of proline around the *S/T*-site. However, indicates that the likelihood of *O*-glycosylation tends to be more favorable in the beta hairpins and beta bridges of proteins. In *ASA_zero* is significant. This appears to be "unusual" because only 2 sequences in the LPM estimation data, both of which are experimentally deemed to be *O*-GlcNAc glycosylated, have ASA values of 0. *ASA_zero* has a point of contact with the concept of an "observation-specific" dummy variable [70-73]. The two sequences with ASA values of 0 are shown in Table 13. An important biological question is whether sequences not exposed at all, but with other *O*-glycosylation friendly characteristics, have a greater propensity to be *O*-glycosylated.

**Table 13: Sequences in the LPM estimation dataset for which ASA values are zero**

| UniProt Accession No. | UniProt Position | Protein Length | Sequence | Secondary Structure | Phi angle | Psi angle | PDB ID | LPM prediction |
|---|---|---|---|---|---|---|---|---|
| P02730 | 162 | 911 | ELLRALLLKH**S**HAGELEALGG | Strand | -68.8 | 160.7 | 1HYN | 89.3% |
| P32119 | 112 | 198 | PLLADVTRRL**S**EDYGVLKTDE | Helix | -60.1 | 40.0 | 1QMV | 110.8% |

The two proteins Peroxiredoxin-2 (P32119) and Band 3 anion transport protein (P02730) have their *O*-linked glycosylation sites buried as indicated by their zero ASA values. While an explanation for this is not immediately apparent, position 112 in Peroxiredoxin-2 (P32119) is a phosphorylation site as well. A question is whether the site is buried in order to make itself amenable to switch between the two processes via a conformational change (*O*-linked and Phosphorylation). Further research is needed to answer this question.

*ASA* is not significant in the LPM. As the two data subsets of sequences in the LPM estimation dataset are either *O*- or *N*- glycosylated, it appears that differences in *ASA*, between *N*- and *O*- glycosylated sequences does not rise to the level of a predictor variable in the fitted LPM. The empirical cumulative probability distribution functions (CDFs) for *ASA*, by estimation data subset, is shown in Figure 2. The mean/median ASA values, shown in Figure 2, differ by LPM estimation data subset. In particular, it appears that ASA values for *N*-glycosylated sequences are generally "larger" than those for *O*-GlcNAc glycosylated sequences. Interestingly, the Wilcoxon rank sum test [41] indicates that the ASA value for *N*-glycosylated sequences is "stochastically larger" [74] than the ASA value for *O*-GlcNAc glycosylated sequences. The *p*-value of this test is about 3.6%. If the KS test is applied, the *p*-value drops to nearly zero (3.96e-13). However, because the first *p*-value is less than 5%, it would be of interest to retest this phenomenon with larger data to see if one can get significantly lower *p*-values, say less than 0.5% [75] with the Wilcoxon rank sum test.

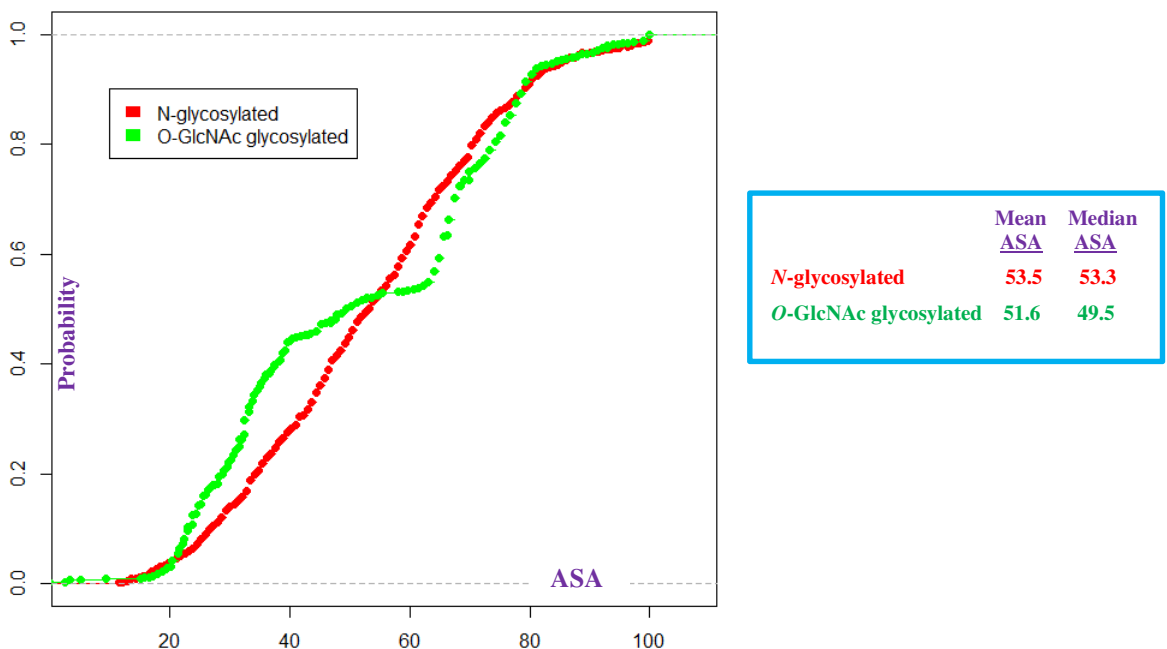

| | Mean ASA | Median ASA |
|---|---|---|
| *N*-glycosylated | 53.5 | 53.3 |
| *O*-GlcNAc glycosylated | 51.6 | 49.5 |

**Figure 2:** Empirical CDFs of ASA values

### *LPM predictions using only sequence data*

The LPM in Table 14 employs only sequence data. The optimal value of *k* for which all predictions are in the 0-1 range is 1.4357. With this *k*, the LPM in Table 14 is estimated by WLS and produces a KS statistic of about 80.4%, which is a steep drop from 99.2% and underscores the germaneness of structural information. Absent 3D structure, this loss in LPM predictive power is not surprising, considering that the same site is used by two fundamental PTM processes driven by conformational flexibility. Maximum separation occurs at the predicted probability value of about 49%. The DW statistic value is about 1.12, indicating, as expected, missing explanatory variables (e.g., structural information) in the LPM as specified. This LPM produced only one prediction outside the 0-1 range flagging a protein deemed to be *O*-glycosylated in the data with a negative (or zero) "probability" (≈ –7.2%) of *O*-glycosylation; its UniProt Accession No., UniProt position, protein length, and sequence are Q16566, 58, 473, and VESELGRGATSIVYRCKQKGT, respectively. The LPM in Table 14 mispredicts (using 50% as the cutoff probability) 54 of the 259 (or 20.85%) out-of-sample *O*-GlcNAc glycosylated sequences as not being *O*-GlcNAc glycosylated. The 95% confidence interval [48] around 20.85% is: (16.35%, 26.20%). If the LPM in Table 14 is applied to a sample of *O*-GalNAc glycosylated sequences, it mispredicts 656 of the 2,079 (or 31.6%) as not being *O*-GalNAc glycosylated. A LPM specific to *O*-GalNAc glycosylation will decrease the mispredictions rate. If the LPM in Table 14 is applied to the *WSTW-Uniprot* dataset (by ignoring the amino acids on either side of the *S/T*-site), 52 of the 236 (≈ 22%) sequences are mispredicted as being *O*-glycosylated. As expected, this is slightly more than 20.85%.

**Table 14: WLS estimated LPM for predicting the probability of *O*-glycosylation given only sequence data**

| *Variable** | *β* | \|*t*\| | *Variable** | *β* | \|*t*\| |
|---|---|---|---|---|---|
| intercept | 0.1379 | 3.27 | p1S | 0.2446 | 5.33 |
| m1C | -0.2020 | 2.11 | p1T | 0.2268 | 4.48 |
| m1F | -0.1373 | 1.96 | p2A | 0.2120 | 4.44 |
| m1P | 0.4708 | 5.44 | p2D | 0.2029 | 2.73 |
| m1T | 0.1535 | 3.25 | p2F | -0.2292 | 2.43 |
| m1V | 0.1348 | 3.41 | p2G | 0.1706 | 3.01 |
| m3F | -0.1950 | 2.92 | p2L | -0.1342 | 2.64 |
| m3L | -0.1367 | 2.65 | p2S | 0.2068 | 4.69 |
| m3P | 0.1852 | 4 | p3Q | 0.1914 | 3.12 |
| m3V | 0.1103 | 2.3 | p3S | 0.1041 | 2.29 |
| m4C | -0.6595 | 4.9 | p4A | 0.2149 | 3.66 |
| m4F | -0.2551 | 4.1 | p4G | 0.2195 | 3.93 |
| m4L | -0.1180 | 2.28 | p4S | 0.1277 | 3.07 |
| m4Q | -0.2802 | 4.52 | p4T | 0.1867 | 3.81 |
| m4V | -0.2253 | 4.37 | p5A | 0.0934 | 1.77 |
| m4Y | -0.2037 | 3.03 | p5C | -0.3927 | 4.63 |
| m5A | 0.1224 | 2.28 | p5L | -0.1029 | 2.04 |
| m5C | -0.2515 | 2.6 | p5N | -0.3141 | 3.6 |
| m5H | -0.3195 | 3.45 | p5W | -0.7183 | 3.72 |
| m5N | -0.2492 | 2.9 | p6C | -0.3787 | 4.2 |
| m5R | 0.1419 | 2.57 | p7A | 0.1763 | 2.71 |
| m5W | -0.2752 | 1.96 | p7C | -0.2791 | 3.06 |
| m5Y | -0.1507 | 2.15 | p7G | 0.1366 | 2.5 |
| m6R | 0.1362 | 2.31 | p7K | 0.1578 | 2.73 |
| m6Y | -0.2011 | 2.49 | p7T | 0.1968 | 3.98 |
| m7L | -0.0910 | 1.86 | p8F | -0.1867 | 2.86 |
| m7M | -0.2490 | 2.83 | p8S | 0.1625 | 3.5 |
| m7R | 0.2907 | 4.38 | pos | 0.2406 | 4.35 |
| m8A | 0.1563 | 2.7 | | | |
| m8K | 0.1726 | 2.9 | | | |
| p1A | 0.2092 | 3.78 | | | |
| p1G | 0.1706 | 3.23 | | | |
| p1K | 0.2524 | 3.37 | | | |
| p1P | 0.4464 | 3.26 | | | |
| p1Q | 0.1728 | 3.05 | | | |

## Concluding Remarks

In order to clarify why the LPM methodology used herein is appropriate, we discuss several points in a question-answer format, in this section.

<u>*Is the LPM a valid methodology to model a binary response?*</u>

The short answer is "yes". However, a longer answer is needed because it is quite unfashionable to use the LPM these days. And that perception needs to be addressed.

The LPM is fairly well-known to econometricians and statisticians [28], [76-81] and has sparked a fairly substantial literature. The LPM is $Y = Xb + u$, but where $Y$ only takes the values 0 and 1. If $E(u) = 0$, then each $u_i$ has variance $E(Y_i)(1 - E(Y_i))$. Goldberger [28] suggested estimating $E(Yi)$ by ordinary LS (OLS), and then re-estimating the model by WLS to achieve homoscedasticity (i.e., constant residual variance) with $E(Y_i)(1 - E(Y_i))$ as the "weights". The primary attraction of the LPM is its simplicity. In large samples, the non-normality of $u$ does not matter. The LPM produces consistent coefficients [32]. And the problem of getting predictions that lie outside the 0-1 range is not an asymptotic one [33]. Yes, "probabilities" less than 0 or greater than 1 can hinder empirical work with finite samples; but, only if those probabilities, per se, are used for something. In our case, probabilities less than 0 are interpreted as nearly 0 and those greater than 1 as nearly 1. This is so, because we are only interested in whether or not a sequence is *O*-glycosylated by assessing whether or not its probability of *O*-glycosylation is greater or less than 50%, respectively. The numerical value, per se, of the estimated probability of *O*-glycosylation below or above this threshold does not matter to us.

Rounding LPM predictions that are less than 0 to a small number close to 0 (e.g., 0.001), and those greater than 1 to a number close to 1 (e.g., 0.999), is a well-known method to constrain the LPM predictions within the 0-1 range. In another approach, the sum of squared errors is minimized subject to the constraints $0 \leq Xb \leq 1$ [82]. OLS LPM predictions are used to reestimate the LPM using WLS to overcome the problem of heteroscedasticity. Another method uses the absolute values of the weights to do WLS estimation. Another method uses only OLS predictions that are in the 0-1 range to do WLS estimation [83]. In another method, the weights are bounded and negative weights are assigned a constant value [84]. Another method generalizes the last two methods ([85], [84], [86]). A limitation of these methods is that their sampling properties are not available. So, we use RR to construct the weights [29].

To avoid the problem, in finite samples, of LPM estimates falling outside of the 0-1 range, logistic regression is a greatly favored alternative to the LPM [87]. The logit model mathematically forces the estimates of $Y$ to lie in the 0-1 range by the transformation: $log(Y \div (1 - Y)) = Xb + u$; and then estimates $Y$. In contrast, an alternative transformation is to use RR to force the estimates of $Y$ in the 0-1 range. However, there is no *a priori* reason that says one way of constraining the estimates of $Y$ to behave like "probabilities" is "superior" to the other. Furthermore, given that most of the predictors in our model (*EQ1*) are also binary variables, it is not at all clear that Nature prefers the likelihood of *O*- glycosylation to vary along an ogive- or sigmoid- like surface, rather than linearly. Nature's preference is even less clear when most of the predictors, as in our case, are 0-1 dummies.

This naturally leads to the question of whether our fitted model satisfies the assumptions of a linear model. For the LPM, the error (residual) term is never normal; rather it is binomial. Furthermore, the error term is not observable – we only observe the final binary outcome in Nature. For large samples the central limit theorem guarantees that LPM errors converge to a normal distribution [88]. Although LPM errors are not observable, one way to *approximate* them is to simply compute them like in ordinary linear regression with a continuous dependent variable – i.e., take the difference of the binary dependent variable, *Y*, and the corresponding LPM prediction; and compute the average squared error, which is the Brier Score. For our model in Table 11, Figure 3 indicates that our sample size is approximately normal with a concentration of small errors in the range –0.04 to +0.04. A normal distribution with a mean of about 0.0025 (which is close to 0) and a standard deviation of about 0.1 fits Figure 3 approximately.

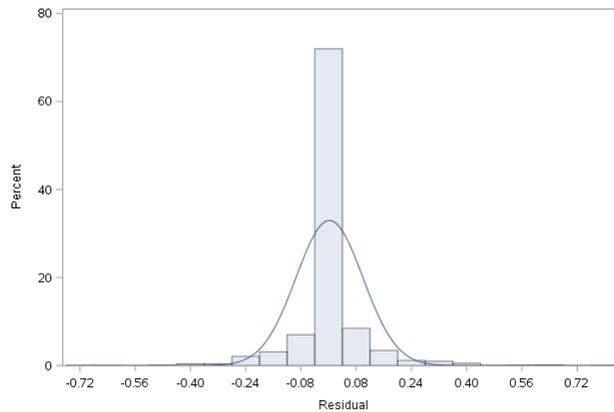

**Figure 3: Distribution of (unobservable) errors for the LPM in Table 11**

Cook's distance is used to test for the presence of outliers [89, 90]. We use the rule of thumb that Cook's distances (*D*s) greater than 0.5 may be problematic. Cook's *D* are shown in Figure 4, which indicates that the assumption of no outliers is a satisfactory one.

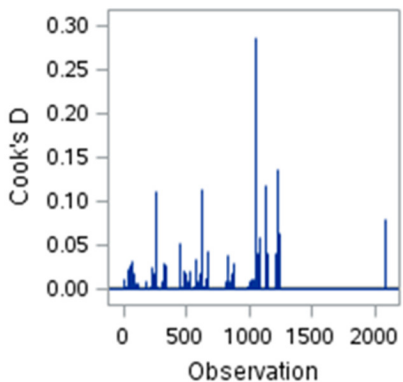

**Figure 4: Cook's distances for the LPM in Table 11**

In the LPM, a unit increase in an independent variable, holding the other independent variables at fixed values, leads to a constant increase in the dependent variable. Because most of our independent variables in the LPM are binary, this is indeed the case. However, for the two continuous variables, the question is whether this constant "partial derivative" or "marginal" effect of an independent variable on the dependent variable holds. In a logit model, for example, this marginal effect varies with the independent variable. A plot of the residuals versus the two continuous variables in our LPM in Table 11 is shown in Figure 5. This plot indicates fairly random scatters. Thus, the constant linear marginal effect assumption is satisfactory.

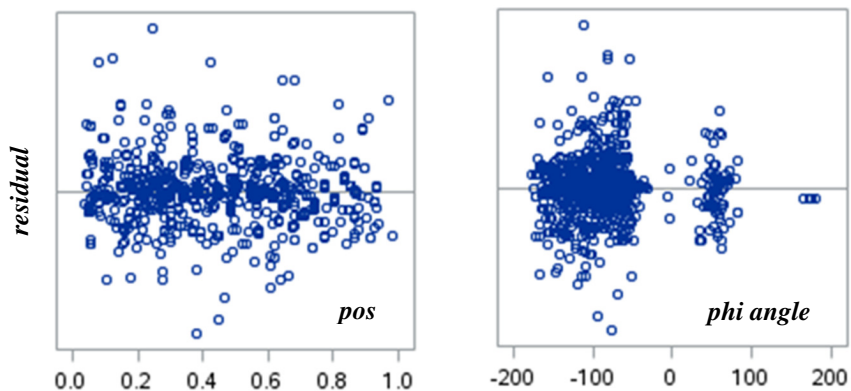

**Figure 5: LPM residuals vs. *pos* and *phi angle***

The LPM is a heteroscedastic model by construction. We first used White's procedure [58] to correct for it. This LS estimated LPM with its heteroscedasticity consistent |*t*| values is shown in Table 15.

**Ridge Regression Estimated Linear Probability Model Predictions of *O*-glycosylation in Proteins with Structural and Sequence Data**

**Table 15**: LS estimated LPM predicting *O*-glycosylation probabilities given structural and sequence data

| Variable* | β | White's $|t|$ | Variable* | β | White's $|t|$ | Variable* | β | White's $|t|$ |
|---|---|---|---|---|---|---|---|---|
| intercept | -0.1177 | 5.17 | m8N | -0.0842 | 3.18 | p7T | -0.1371 | 6.97 |
| m1D | 0.1827 | 5.30 | m8P | -0.1387 | 6.82 | p7V | -0.1173 | 4.68 |
| m1L | 0.0930 | 3.71 | m8V | -0.2103 | 7.60 | p7Y | 0.1188 | 4.05 |
| m1P | 0.7844 | 13.86 | p1A | 0.1754 | 5.59 | p8H | 0.1194 | 2.51 |
| m1R | -0.0619 | 2.34 | p1D | 0.0984 | 2.65 | p8K | 0.0484 | 1.52 |
| m3A | 0.2046 | 8.22 | p1F | 0.0831 | 2.43 | p8N | 0.1052 | 3.22 |
| m3C | -0.1245 | 4.37 | p1S | 0.0884 | 3.27 | p8Q | 0.1466 | 4.52 |
| m3L | -0.1099 | 6.18 | p1T | 0.1567 | 4.59 | pos | 0.2509 | 8.51 |
| m3N | -0.2151 | 3.23 | p1V | 0.0623 | 3.34 | ASA_zero | 0.7952 | 9.89 |
| m3T | -0.0743 | 2.90 | p2A | 0.0697 | 1.33 | II | 0.1387 | 3.12 |
| m4E | 0.1109 | 2.61 | p2H | 0.2885 | 4.65 | II′ | -0.2684 | 4.04 |
| m4F | -0.0881 | 3.43 | p2P | 0.2730 | 9.84 | Helix | 0.1508 | 7.66 |
| m4N | 0.1405 | 3.27 | p2Q | -0.1379 | 4.70 | Beta Bridges | 0.6330 | 9.35 |
| m4R | 0.1408 | 3.71 | p2Y | -0.1906 | 3.98 | Beta Hairpin | 0.8016 | 13.87 |
| m4V | -0.0560 | 2.91 | p3N | -0.1483 | 5.13 | Beta Hairpin Strand | -0.0916 | 5.34 |
| m5D | 0.2333 | 6.89 | p3W | -0.1152 | 3.37 | Phi angle | -0.0003 | 3.04 |
| m5F | -0.0941 | 2.85 | p4A | 0.1650 | 4.77 | | | |
| m5G | 0.0560 | 1.75 | p4E | 0.1711 | 6.75 | | | |
| m5I | -0.0872 | 2.61 | p4P | 0.1946 | 8.26 | | | |
| m5Y | -0.1326 | 4.60 | p5C | -0.1742 | 5.23 | | | |
| m6E | 0.0916 | 3.22 | p5E | 0.1512 | 4.93 | | | |
| m6H | -0.1270 | 5.34 | p5H | 0.2173 | 4.08 | | | |
| m6V | 0.1567 | 3.99 | p5I | -0.0783 | 3.52 | | | |
| m6W | 0.1063 | 2.81 | p5Q | 0.1209 | 3.16 | | | |
| m6Y | -0.1267 | 4.11 | p5T | 0.0651 | 2.79 | | | |
| m7A | 0.1622 | 5.51 | p5Y | -0.1818 | 4.60 | | | |
| m7E | -0.1072 | 4.70 | p6F | -0.0883 | 1.97 | | | |
| m7G | -0.0703 | 3.42 | p6G | -0.1561 | 5.14 | | | |
| m7H | 0.2640 | 5.12 | p6M | 0.3066 | 4.87 | | | |
| m7I | 0.1141 | 5.62 | p6N | -0.1144 | 5.29 | | | |
| m7K | 0.1366 | 3.55 | p6Q | -0.2447 | 7.92 | | | |
| m7S | 0.0859 | 2.72 | p7A | 0.1803 | 3.10 | | | |
| m8G | 0.0964 | 4.51 | p7C | -0.1207 | 3.99 | | | |
| m8L | -0.1518 | 5.28 | p7E | -0.1148 | 4.67 | | | |
| m8M | -0.1187 | 4.49 | p7G | -0.0945 | 3.50 | | | |

Because the form of the heteroscedasticity in the LPM is known, we had also used WLS to correct for it. This was done by estimating *Y* by RR in such a manner so as to guarantee that the predicted *Y* lies in the unit interval [29]. The plots of the actual observations (which are binary) vs. the RR predictions and the final WLS predictions of our model in Table 11 are shown in the left-hand and right-hand panels of Figure 6, respectively.

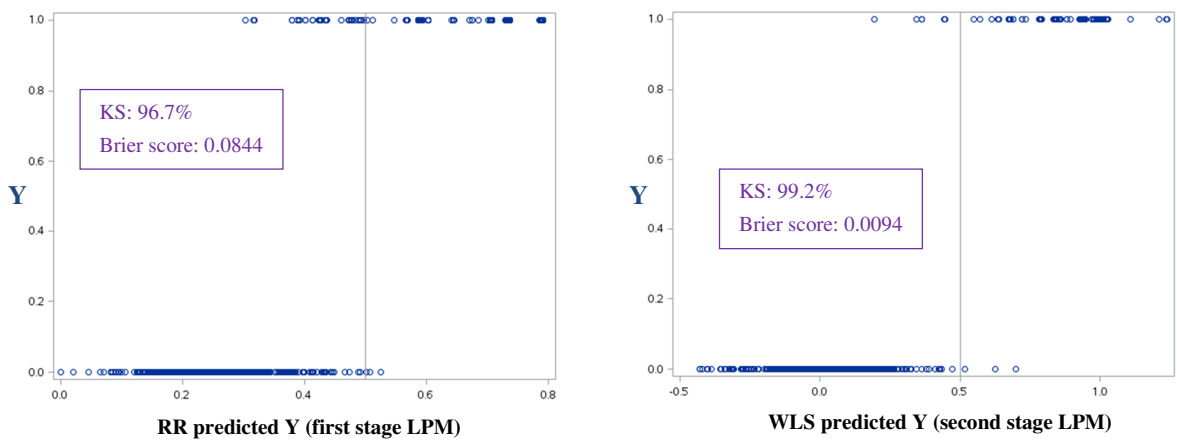

**Figure 6: Actual vs. predicted outcomes of the LPM in Table 11**

The RR predicted $Y$ in the left-hand panel of Figure 6 was used to estimate the LPM residual variances (this is the "first stage" LPM), so the problem of non-constant variances in the LPM could be corrected for by using WLS. As can be visually observed in the left-hand panel, all predictions are constrained to lie in the 0-1 range for the smallest value of $k$ [29]. That is, similar to logistic regression, a mathematical constraint has been imposed on the predicted $Y$. In logistic regression this is done via a transformation; here it is done by constraining the estimated coefficients. It is also known [30, 31] that the RR predicted LPM (in the left-hand panel of Figure 6) is also a competitive model. That is, one does not necessarily have to compute the predictions in the right-hand panel of Figure 6, because RR has the capability to account for non-constant residual variances [30]. The thing to note is that the KS drops and the Brier score increases in the left-hand panel of Figure 6 relative to the corresponding values in the right-hand panel [57]. This may be interpretable as the "cost" for constraining the predicted $Y$ to lie in the 0-1 range, rather than leveraging the asymptotic property that predictions outside the 0-1 range do not matter in large samples.

### *What is the RR estimated LPM ANOVA?*

The RR estimated LPM that generated Figure 6 is shown in Table 16. The RR estimated standard errors for the coefficients are also shown in Table 16. Because RR produces non-integer degrees of freedom, unlike in classical LS estimation, the resultant RR analysis of variance (ANOVA) table is shown in Table 17 [91]. Note that the RR estimated mean square error (MSE) is about 0.0827, which is very much like the Brier Score in the left-hand panel of Figure 6. RR produces an approximate, but not exact, $F$-ratio (for non-stochastic $k$) in the ANOVA table in Table 17, which is about 341 [92]. The angle between the regression vector and the residual vector is about 37º, which measures "how much the residual vector deviates from being orthogonal to the $X$ space as it is in least squares" [95] [93].

**Table 16**: RR estimated LPM predicting *O*-glycosylation probabilities given structural and sequence data

| Variable | $\beta$ | Standard error | Variable | $\beta$ | Standard error | Variable | $\beta$ | Standard error |
|---|---|---|---|---|---|---|---|---|
| intercept | 0.3124 | 0.0322 | m8N | -0.0595 | 0.0496 | p7T | -0.0660 | 0.0364 |
| m1D | 0.0343 | 0.0415 | m8P | -0.0638 | 0.0336 | p7V | -0.0556 | 0.0386 |
| m1L | 0.0560 | 0.0307 | m8V | -0.0687 | 0.0433 | p7Y | 0.0463 | 0.0439 |
| m1P | 0.1345 | 0.0687 | p1A | -0.0119 | 0.0435 | p8H | 0.0487 | 0.0459 |
| m1R | -0.0671 | 0.0391 | p1D | 0.0614 | 0.0474 | p8K | 0.0037 | 0.0409 |
| m3A | 0.0127 | 0.0368 | p1F | 0.0657 | 0.0396 | p8N | 0.0638 | 0.0428 |
| m3C | -0.0633 | 0.0473 | p1S | -0.0007 | 0.0383 | p8Q | 0.0082 | 0.0414 |
| m3L | 0.0228 | 0.0316 | p1T | 0.0664 | 0.0391 | pos | 0.0338 | 0.0417 |
| m3N | -0.0528 | 0.0703 | p1V | -0.0329 | 0.0323 | ASA_zero | 0.1204 | 0.0690 |
| m3T | -0.0515 | 0.0406 | p2A | 0.0554 | 0.0465 | II | -0.0266 | 0.0547 |
| m4E | -0.0219 | 0.0543 | p2H | 0.0169 | 0.0592 | II´ | -0.0616 | 0.0678 |
| m4F | 0.0251 | 0.0377 | p2P | 0.0788 | 0.0349 | Helix | 0.0603 | 0.0271 |
| m4N | 0.0548 | 0.0464 | p2Q | -0.0658 | 0.0444 | Beta Bridges | 0.1283 | 0.0588 |
| m4R | 0.0669 | 0.0476 | p2Y | -0.0701 | 0.0602 | Beta Hairpin | 0.1353 | 0.0689 |
| m4V | -0.0619 | 0.0309 | p3N | -0.0574 | 0.0469 | Beta Hairpin Strand | -0.0641 | 0.0264 |
| m5D | 0.0553 | 0.0444 | p3W | -0.0584 | 0.0619 | Phi angle | 0.0001 | 0.0002 |
| m5F | -0.0751 | 0.0465 | p4A | 0.0642 | 0.0406 | | | |
| m5G | -0.0351 | 0.0482 | p4E | 0.0139 | 0.0350 | | | |
| m5I | -0.0408 | 0.0478 | p4P | 0.0384 | 0.0359 | | | |
| m5Y | -0.0577 | 0.0482 | p5C | -0.0702 | 0.0473 | | | |
| m6E | 0.0249 | 0.0338 | p5E | -0.0222 | 0.0471 | | | |
| m6H | -0.0693 | 0.0579 | p5H | 0.0701 | 0.0540 | | | |
| m6V | 0.0545 | 0.0395 | p5I | -0.0552 | 0.0419 | | | |
| m6W | 0.0067 | 0.0525 | p5Q | 0.0310 | 0.0511 | | | |
| m6Y | -0.0589 | 0.0450 | p5T | -0.0380 | 0.0371 | | | |
| m7A | 0.0418 | 0.0381 | p5Y | -0.0697 | 0.0546 | | | |
| m7E | -0.0655 | 0.0398 | p6F | -0.0644 | 0.0482 | | | |
| m7G | -0.0569 | 0.0348 | p6G | -0.0540 | 0.0418 | | | |
| m7H | 0.0707 | 0.0491 | p6M | 0.0608 | 0.0620 | | | |
| m7I | -0.0053 | 0.0385 | p6N | -0.0695 | 0.0438 | | | |
| m7K | 0.0396 | 0.0409 | p6Q | -0.0782 | 0.0484 | | | |
| m7S | -0.0403 | 0.0433 | p7A | 0.0765 | 0.0562 | | | |
| m8G | 0.0512 | 0.0390 | p7C | -0.0576 | 0.0493 | | | |
| m8L | -0.0587 | 0.0351 | p7E | -0.0513 | 0.0393 | | | |
| m8M | -0.0693 | 0.0525 | p7G | 0.0306 | 0.0383 | | | |

**Table 17**: RR estimated LPM ANOVA with $k = 3.77895$

| Source | Degrees of Freedom | Sum of Squares | Mean Square | Approximate $F$-Ratio | Angle |
|---|---|---|---|---|---|
| Regression | 4.2906 | 121.1983 | 28.2472 | | |
| Residual | 2043.1994 | 169.0489 | 0.0827 | 341.4085 | 37.82º |
| Nonorthogonal component (NON) | 21.5099 | 226.1397 | | | |
| Total | 2069 | 516.387 | | | |

*Is the lasso penalized logit model (LPLM) more appropriate than the LPM to estimate Y?*

The concept of the "lasso" penalty to model a dataset in which the number of predictors exceeds the number of observations goes back, at least, to the 1970s [94-96]. Later it was made fashionable by describing its applicability to statistical model selection [96].

The LPLM is estimated by minimizing the negative log-likelihood function, $L(\cdot)$, subject to the constraint that the sum of the absolute values of the coefficients is less than or equal to some value – that is, $||b||_1 \leq t$, where $log[(Pr(Y=1) \div (1 - Pr(Y=1))] = Xb$ and $||v||_p$ denotes the "*p*-norm" of a vector *v*. In contrast, RR uses the constraint $(||b||_2)^2 \leq t$. That is, the sum of squared coefficients, instead of absolute coefficients, is constrained. In Lagrangian form [97, 98], the LPLM optimization problem can be recast as: minimize $-L(\cdot) + \lambda \times sum(|b|)$. This minimization is done using Nesterov's method [99]. Here, a regularization parameter, $\rho^i$, where $0 < \rho < 1$ and $i$ is the $i^{th}$ step in Lasso selection, is used for $\lambda$ in the $i^{th}$ step. Variable subset selection, via the Lasso, is done, using up to 200 steps, so as to minimize the Schwarz Bayesian Criterion (SBC) [100]. In sample, several LPLMs are estimated for different values of $\rho$ and the resultant number of variables (excluding the intercept) selected, the KS statistic values, the Brier Scores, and twice the negative log-likelihood values are shown in Table 18. In terms of the SBC, the "optimal" LPLM, on the full sample, is the one with $\rho = 0.815$; and this is also nearly "optimal" in terms of Brier Score and KS statistic value. To assess LPLM performance out of sample as $\rho$ varies, 5-fold cross validation (CV) is done. LPLMs are fit on a set of 4 "training" folds and the Brier Score is calculated on the holdout fold. Thus, for a particular choice of $\rho$, five out of sample Brier Scores are available. The signal-to-noise ratio (SNR), which is $\mu/\sigma$, of these 5 Brier Scores is calculated. Per the *Rose criterion* [101], SNRs above 5 are deemed to be "good" in the engineering literature. The SNRs are shown in the last row of Table 18; and at $\rho$ of 0.82, the SNR is a maximum and satisfies the Rose criterion. Thus, overall, it is reasonable to take $\rho = 0.82$ as "optimal".

**Table 18**: Summary of estimated LPLMs as $\rho$ varies

| | $\rho$ | 0.05 | 0.1 | 0.2 | 0.3 | 0.4 | 0.5 | 0.6 | 0.7 | 0.8 | 0.81 | 0.815 | 0.82 | 0.85 |
|---|---|---|---|---|---|---|---|---|---|---|---|---|---|---|
| *In sample* | SBC | 1710 | 1318 | 1257 | 1133 | 1036 | 931 | 878 | 798 | 761 | 749 | 737 | 751 | 723 |
| | Variables | 12 | 39 | 92 | 91 | 73 | 34 | 50 | 41 | 38 | 54 | 51 | 51 | 39 |
| | KS (%) | 78.0 | 89.4 | 94.1 | 95.7 | 95.2 | 91.3 | 95.2 | 95.5 | 95.7 | 96.8 | 96.7 | 96.6 | 95.8 |
| | Brier Score | 0.115 | 0.062 | 0.031 | 0.023 | 0.025 | 0.038 | 0.026 | 0.026 | 0.025 | 0.017 | 0.018 | 0.019 | 0.022 |
| | $-2 \times L(\cdot)$ | 1610 | 1013 | 547 | 431 | 471 | 664 | 489 | 478 | 463 | 329 | 340 | 354 | 418 |
| | | | | | | | | | | | | | | |
| *Out of sample* | Brier Score SNR | 1.76 | 2.55 | 1.47 | 1.86 | 2.18 | 1.85 | 3.42 | 1.59 | 4.32 | 3.75 | 2.00 | 7.78 | 3.06 |

The resultant LPLM is shown in Table 19. An important question is: if the LPLM selected variables drive *O*-glycosylation in Nature, would they, post-Lasso, continue to be statistically significant if they were put into a classical logit model to reestimate *Y* ?

**Table 19**: The selected LPLM with $\rho = 0.82$

| *Variable* | $\beta$ | *Variable* | $\beta$ | *Variable* | $\beta$ | *Variable* | $\beta$ |
|---|---|---|---|---|---|---|---|
| intercept | −3.7350 | m6S ☼ | −0.4411 | p2I* | −0.3424 | p7T | −1.0516 |
| m1D* | 0.0982 | m6V | 0.5139 | p2L | −0.0053 | p8G* ☼ | −0.0074 |
| m1L* | 1.0127 | m7H | 2.3444 | p2P | 2.6960 | p8K ☼ | 0.5589 |
| m1R* | −0.1413 | m7K | 0.8149 | p2V* ☼ | 0.0130 | p8N* | 0.5041 |
| m1S | −0.1913 | m7L* ☼ | −0.4651 | p3T* | 0.4597 | p8Q* | 0.3987 |
| m3A* | 0.5288 | m8A | 0.5662 | p4A* | 0.9764 | pos | 0.3384 |
| m3G | 1.1410 | m8V | −0.0763 | p4E | 1.3507 | ASA* ☼ | −0.0144 |
| m4N ☼ | 0.1992 | p1D* | 0.4975 | p4H | −0.1231 | I* | −0.2885 |
| m4R | 0.0413 | p1E* ☼ | −0.3194 | p4I* ☼ | −0.0779 | Helix | 1.7757 |
| m4V* | −0.2451 | p1F* | 0.9501 | p5H | 0.1778 | BH | 0.3832 |
| m4Y* ☼ | −0.0003 | p1L | −0.7827 | p6L | 1.2946 | BH_strand | −0.8759 |
| m5D | 1.1952 | p1S | 0.1261 | p6T* | 1.0123 | Phi angle* | −0.0047 |
| m6E | 0.0586 | p1T | 2.3710 | p7A | 1.4453 | Psi angle* | −0.0011 |

* / ☼  not significant at 10% in the "equivalent" classical logit model / LPM

Using the classical logit model, we reestimated $Y$ with the independent variables in Table 19. The classical logit model could not be estimated because of the problem of "quasi-complete separation", which makes finding the optimal maximum likelihood solution difficult. This problem arises when $Y$ separates a subset of the predictors to a large enough degree. To resolve this problem, we used Firth's penalized maximum likelihood method [101, 102]. Several predictors (23 of the 51) turn out to be statistically insignificant at the 10% level of significance, and these are noted by asterisks in Table 19. Next, $Y$ is reestimated using the LPM with the predictors in Table 19. The predictors that are statistically insignificant (10 of the 51), in terms of White $t$-values, at the 10% level of significance are noted in Table 19 by the sun-like symbol "☼". The LPM also indicates there is no multicollinearity among the predictors, because all of the VIFs are less than 10. However, the DW statistic drops to 0.8, likely indicating there are missing explanatory variables [85].

Of course, given a set of predictors, the estimated coefficients produced by the classical logit model, the LPM, and the LPLM are not comparable, because different optimization procedures are used to estimate $Y$. However, a philosophical question arises. Suppose, Nature operates along a logit surface to reveal the probability of *O*-glycosylation, given sequence and structural information in the neighborhood of the *S/T* site; and the Lasso has identified the relevant predictors. It appears to us that the selected predictors should, for the most part, be statistically significant, in the classical sense, when they are put into a classical logit model. The fact that 23 of the 51 LPLM predictors are not significant is troubling.

The rates at which the estimated models mispredict $Y$ as 1 when it is empirically 0 (i.e., not *O*-glycosylated) in the estimation dataset, and 0 when it is empirically 1 (i.e., *O*-glycosylated), is shown in Table 20, together with the fit statistics. A 50% cutoff probability is used to measure the mispredictions rate.

**Table 20: Mispredictions rate and fit statistics for selected models**

| Model | *In the set of non-O-glycosylated sequences (Y=0), the percentage of those that have estimated probabilities of O-glycosylation greater than 50% ($\hat{Y} > 0.5$)* | *In the set of O-glycosylated sequences (Y=1), the percentage of those that have estimated probabilities of O-glycosylation less than or equal to 50% ($\hat{Y} \leq 0.5$)* | *KS* | *Brier Score* |
|---|---|---|---|---|
| Ordinary LS estimated LPM in Table 15 | 0.37 | 0.61 | 99.1% | 0.009 |
| RR estimated LPM (used for estimating the weights for the WLS estimated LPM in Table 11) | 0.28 | 7.90 | 96.7% | 0.084 |
| LPM in Table 11 | 0.28 | 0.61 | 99.2% | 0.009 |
| LPLM with $\rho = 0.82$ in Table 19 | 0.83 | 3.55 | 96.6% | 0.019 |

As can be seen in Table 20, our WLS estimated LPM (Table 11) is performing quite well and does have an edge over the other models. This is not surprising. For example, the LPLM selects a subset of variables by setting certain coefficients to exactly 0. In our context, the interpretation of coefficients that are exactly zero is not easy. Coefficients of variables may be 0 if they are "irrelevant" and uncorrelated with the "relevant" variables. If many variables are irrelevant one possibility is that Nature has "wasted" certain amino acids by placing them in certain positions along the sequence; and another possibility is that they are "silent", rather than "wasted". In contrast, RR never sets coefficients to exactly 0. Rather, RR exhibits optimal linear shrinkage of the coefficients [103]. In particular, simulation studies indicate that the performance of RR dominates those of its peers such as partial least squares and principal component regression [103]. If we assume that Nature has not "wasted" amino acids, then RR is better suited to modify coefficients because amino acids in certain positions may have a "small" effect that is not visible. That is, the "relevant" amino acids may dominate the "irrelevant" ones (until a "signal" changes that) in determining the three-dimensional structure of the protein, and, hence, its function. Next, we present an example to clarify this point.

Consider the case of the *psi* angle. It is not a predictor in Table 11, but is a predictor in Table 19. Because the *phi* and *psi* angles go together, an important question is whether the LPM in Table 11 failed to capture it as a predictor, while the LPLM in Table 19 captured it. The answer is "no"; because to a significant degree the psi angle is determined by the amino acids and the positions they occupy. To see this, a linear regression is fit with the psi angle as the dependent variable and sequence information as independent variables. The linear regression model is chosen via two steps. In the first step, stepwise variable selection under 5-fold CV is done so as to minimize the CV PRESS. The minimum CV PRESS attained was about 4,253,609 with 106 independent variables in the specification. Then, in the second step, because the model is heteroscedastic, the White *t*-values [58] are used to retain only those variables that have *p*-values less than 10% – i.e., variable subset selection is done, manually, using White *t*-values instead of classical *t*-values. The selected model is in Table 21.

**Table 21: Regressing psi against sequence information**

| Variable | $\beta$ | White's \|t\| | Variable | $\beta$ | White's \|t\| | Variable | $\beta$ | White's \|t\| |
|---|---|---|---|---|---|---|---|---|
| intercept | 164.7095 | 15.76 | m6T | -27.7795 | 2.28 | p4P | -63.4680 | 6.26 |
| m1E | -41.8628 | 2.97 | m7E | 43.9542 | 5.47 | p5C | -73.2859 | 4.80 |
| m1F | 13.3555 | 1.67 | m7K | -32.2923 | 3.01 | p5E | -59.8629 | 6.27 |
| m1G | -39.8688 | 5.61 | m7L | 29.8517 | 3.96 | p5F | -46.9835 | 3.43 |
| m1I | 39.5770 | 4.07 | m7V | -47.2830 | 5.75 | p5H | -103.8328 | 6.47 |
| m1P | -88.6389 | 2.43 | m7Y | -21.9592 | 1.70 | p5K | -46.4349 | 5.08 |
| m1S | 22.2152 | 2.37 | m8G | 26.2222 | 2.74 | p5M | -55.0010 | 2.58 |
| m1V | 22.1718 | 2.53 | m8I | -26.1921 | 1.96 | p5R | -49.2451 | 3.17 |
| m1W | -56.8472 | 1.73 | m8K | 39.6707 | 3.18 | p5T | -26.0844 | 3.06 |
| m3C | -107.3697 | 7.69 | m8L | -59.0523 | 7.31 | p5W | -92.6251 | 6.31 |
| m3H | -34.3322 | 2.59 | m8M | -79.0357 | 3.70 | p6C | 45.0324 | 4.28 |
| m3I | 55.7918 | 3.38 | m8Q | -48.2257 | 3.54 | p6M | -103.3734 | 6.33 |
| m3P | -60.7537 | 6.03 | m8V | 20.4134 | 1.93 | p6R | 42.4221 | 4.59 |
| m3Q | -91.6187 | 6.83 | p1E | 21.3198 | 2.71 | p6S | 33.0045 | 3.96 |
| m3R | -78.6022 | 7.07 | p1I | 57.7033 | 3.69 | p6W | 78.9544 | 2.14 |
| m3S | -58.6653 | 4.59 | p1P | 168.9937 | 4.72 | p7M | -30.9223 | 2.37 |
| m3Y | -43.6250 | 4.75 | p1T | 25.9053 | 2.81 | p7Q | -74.3226 | 5.24 |
| m4A | -22.2813 | 3.06 | p2A | -32.4674 | 2.98 | p7S | -27.8055 | 2.90 |
| m4D | 34.2970 | 2.86 | p2I | -51.3731 | 6.62 | p7V | -25.1988 | 2.55 |
| m4F | -30.2667 | 2.51 | p2Q | 22.8051 | 2.34 | p7W | -126.1256 | 5.32 |
| m4H | 42.5802 | 2.67 | p2R | -73.2310 | 6.60 | p7Y | -33.3375 | 3.33 |
| m4I | 52.7538 | 6.23 | p2T | -24.7715 | 2.04 | p8C | -32.8242 | 2.11 |
| m4S | -17.8569 | 2.04 | p2V | 28.9007 | 3.16 | p8G | 24.3492 | 3.24 |
| m5C | -61.3798 | 5.19 | p3A | -33.2607 | 3.75 | p8K | -26.5478 | 3.16 |
| m5M | -50.2581 | 2.77 | p3L | -32.1617 | 3.65 | p8Q | 30.6688 | 2.49 |
| m5N | -45.0953 | 2.62 | p3Q | 44.6698 | 2.94 | p8S | 47.6755 | 4.08 |
| m5Q | 36.6723 | 4.03 | p3T | -34.4732 | 3.71 | p8V | 24.9615 | 3.24 |
| m5R | 36.6709 | 4.25 | p4A | -22.9135 | 2.38 | p8Y | 29.0752 | 2.56 |
| m6A | -51.1224 | 3.88 | p4C | -54.2233 | 3.33 | pos | -45.3919 | 4.67 |
| m6D | -31.7667 | 2.24 | p4D | 35.4777 | 3.29 | ASA | -0.6246 | 5.84 |
| m6K | -16.0558 | 2.11 | p4E | 53.4754 | 6.01 | | | |
| m6L | -30.2475 | 3.66 | p4F | 41.2684 | 3.29 | | | |
| m6P | -62.2979 | 6.04 | p4G | 19.1452 | 2.11 | | | |
| m6R | -22.4489 | 2.12 | p4L | -21.9163 | 2.72 | | | |
| m6S | 30.1454 | 3.69 | p4N | 31.5863 | 1.94 | | | |

The model in Table 21 has 99 variables, and an adjusted $R^2$ of about 76%. However, its standardized (studentized) residuals are not normally distributed as indicated by the Q-Q plot in Figure 7 [104]. This is not surprising, because *psi* is a "circular" measure and has heavy tails as can be seen by its tendency to cluster (with *phi*) in certain (and remain sparse in other) regions of the X-Y plane, which can be seen in the Ramachandran plot in Figure 8 [105]. The non-normality of the residuals indicates the need to do more work to normalize *psi*. However, as fitting a model to *psi* is not the focus of this paper, we did not pursue that goal at this time (our focus is on answering the question raised in the last paragraph). Nonetheless, the *t*-values can be viewed as approximations (e.g., SNRs) for model selection by viewing linear regression, in this case, as purely an optimization method. In particular, we held the 99 variables in Table 21 "constant" and redid 5-fold CV 50 times to study the variability in out of sample CV PRESS values. We did this because we noted that in the second step used to select the model (based on White's *t*-values), the new CVPRESS tends to exceed the minimum CV PRESS of about 4,253,609 – the reduction of 106 variables to 99 variables is fairly small and likely explains this movement. This variability, with its associated summary statistics, is shown in Figure 9 by plotting the percentage increase in each new CV PRESS over the minimal CV PRESS (i.e., 4,253,609). Figure 9 indicates that this variation is about 4.5% on average, which is not large enough to reject the assumption that the distribution of amino acids along the sequence plays a significant role in determining structure and, hence, function. Future research can more precisely mathematize this role.

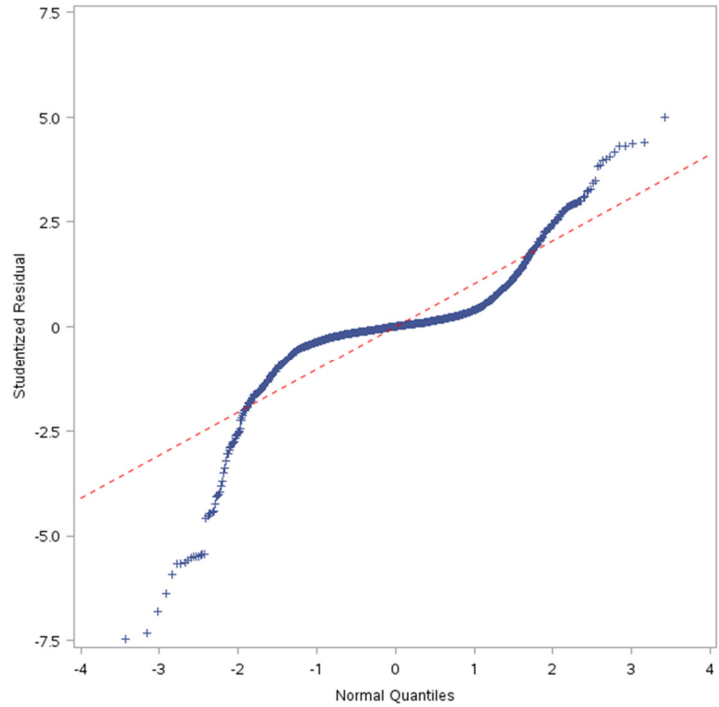

**Figure 7**: Q-Q plot of residuals for the model in **Table 21**

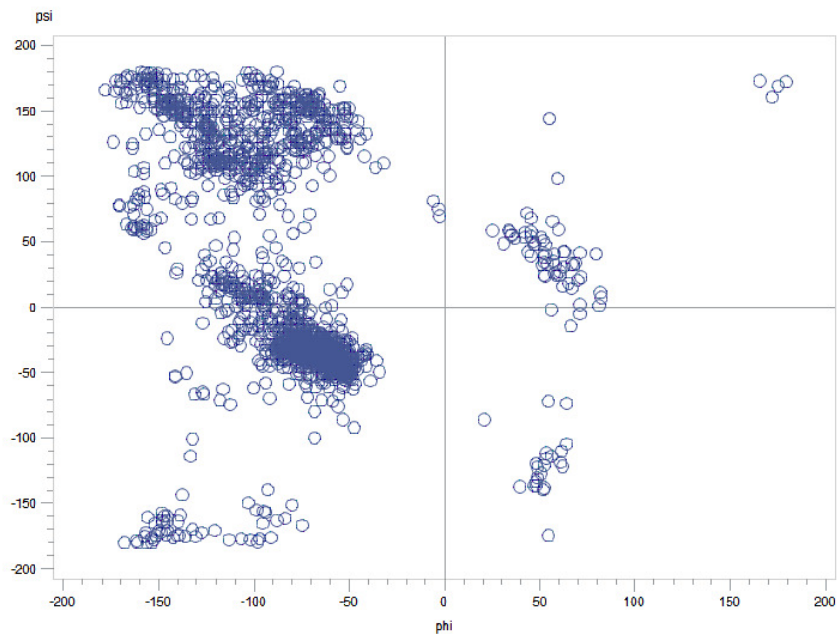

**Figure 8**: Ramachandran Plot

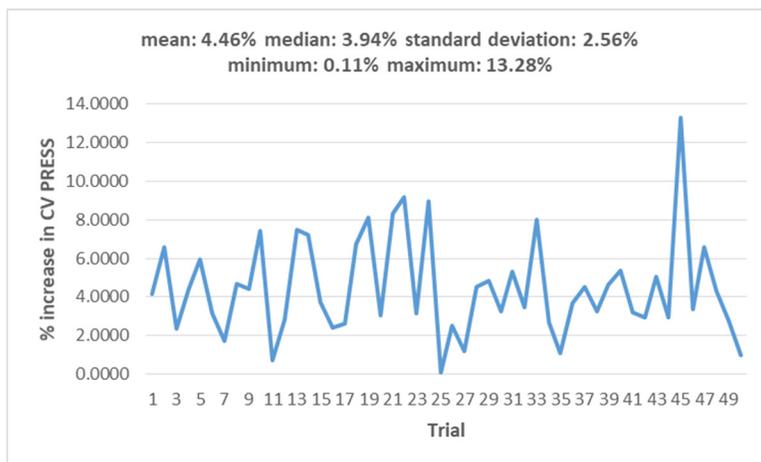

<span style="color:#2E9FD6">**Figure 9**</span>: **Increases in CV PRESS when the model in** <span style="color:#2E9FD6">**Table 21**</span> **is subject to 5-fold CV 50 times**

The number of times predictors are selected by the LPLM as $\rho$ varies in <span style="color:#2E9FD6">Table 18</span> is shown in <span style="color:#2E9FD6">Table 22</span>. In particular, $\rho$ takes 13 values in <span style="color:#2E9FD6">Table 18</span>. For each $\rho$, an optimal LPLM is selected. In all, there are 13 optimal LPLMs. All possible variables are shown in a rectangular format in <span style="color:#2E9FD6">Table 22</span>. The positions of amino acids along the sequence is shown in an "integer line" format in the penultimate row of <span style="color:#2E9FD6">Table 22</span>. So, for example, if we are at position number 3 to the right of *S/T* in <span style="color:#2E9FD6">Table 22</span> and go up vertically 5 "cells" we would arrive at the row for amino acid *G*; and this cell would represent the dummy variable *p3G* – that is, the dummy variable representing amino acid *G* in position +3. The variables (predictors) in the LPM of <span style="color:#2E9FD6">Table 11</span> are shaded in green in <span style="color:#2E9FD6">Table 22</span>. For the 13 LPLMs, the frequency with which a predictor occurs in the LPLMs is noted in <span style="color:#2E9FD6">Table 22</span>. For example, the number 12 in cell (–3, G) in <span style="color:#2E9FD6">Table 22</span> means that 12 of the 13 LPLMs had *m3G* as a predictor. Similarly, the 13 in cell (8, N) means that *p8N* was selected as a predictor by all of the 13 LPLMs; and, additionally, because this cell is shaded green, *p8N* is also a predictor in the selected LPM of <span style="color:#2E9FD6">Table 11</span>. As another contrasting example, the predictor *BH* (Beta Hairpin) was only selected in 3 of the 13 LPLMs, but is in the LPM of <span style="color:#2E9FD6">Table 11</span>. In summary, the wide variation in subsets of selected LPLM predictors, as $\rho$ varies, is captured in <span style="color:#2E9FD6">Table 22</span>.

**Ridge Regression Estimated Linear Probability Model Predictions of *O*-glycosylation in Proteins with Structural and Sequence Data**

**Table 22: Frequencies of variables selected by the LPLM as $\rho$ varies***

| | | -8 | -7 | -6 | -5 | -4 | -3 | -2 | -1 | S/T | 1 | 2 | 3 | 4 | 5 | 6 | 7 | 8 |
|---|---|---|---|---|---|---|---|---|---|---|---|---|---|---|---|---|---|---|
| **STRUCTURAL** | **BH Strand** | | | | | | | 12 | | | | | | | | | | |
| | **Strand** | | | | | | | 3 | | | | | | | | | | |
| | **Beta Hairpin (BH)** | | | | | | | 3 | | | | | | | | | | |
| | **Beta Bridges** | | | | | | | | | | | | | | | | | |
| | **VIII** | | | | | | | 3 | | | | | | | | | | |
| | **IV** | | | | | | | 2 | | | | | | | | | | |
| | **II′** | | | | | | | | | | | | | | | | | |
| | **II** | | | | | | | | | | | | | | | | | |
| **VARIABLE** | **I** | | | | | | | 13 | | | | | | | | | | |
| | **helix** | | | | | | | 12 | | | | | | | | | | |
| | **asa = 0** | | | | Observation-specific dummy is not estimable in a classical logit model [71, 72] | | | | | | | | | | | | | |
| | **asa** | | | | | | | 13 | | | | | | | | | | |
| | **psi** | | | | | | | 13 | | | | | | | | | | |
| | **phi** | | | | | | | 13 | | | | | | | | | | |
| | **pos** | | | | | | | 5 | | | | | | | | | | |
| **AMINO** | **Y** | | | 3 | | 7 | 2 | | | | | | | | | | 1 | |
| | **W** | | 4 | | | | 1 | | | | | | | | | | | |
| | **V** | | 7 | 11 | 1 | 12 | | | | | 12 | 3 | | | | 12 | 11 | |
| **ACIDS** | **T** | | | | | | | | | | 12 | | 12 | | | | 11 | |
| | **S** | | 3 | 12 | | | | | 6 | | 6 | 4 | | 3 | | | 1 | |
| | **R** | | | | | 7 | 2 | | 11 | | | | | | | | 1 | |
| **as** | **Q** | | | | 2 | | | | | | | | | | 5 | 2 | | 10 |
| | **P** | 2 | 3 | | | | | | 13 | | | 13 | | 3 | 1 | 1 | | |
| | **N** | | | 1 | 3 | 10 | | | | | | 3 | | | | | 2 | 13 |
| **DUMMIES** | **M** | | | | | 1 | 4 | | | | | | | | | 2 | | |
| | **L** | 3 | 12 | 3 | 4 | 8 | | | 13 | | 12 | 3 | | | 2 | 12 | 1 | |
| | **K** | 1 | 10 | 1 | | 4 | 1 | | | | | | | | | | | 10 |
| | **I** | | | 5 | | | | | 1 | | | 12 | 3 | 7 | | | | 10 |
| | **H** | | 12 | | | | | | | | | | | 6 | 12 | | | 3 |
| | **G** | 4 | | | | | 13 | | | | | 5 | | | | | | 10 |
| | **F** | 1 | | | | | | | | | 13 | | | | 1 | | | |
| | **E** | 4 | | 5 | | | 1 | | | | 12 | | | 11 | | | | 3 |
| | **D** | | | | 12 | | | | 6 | | 11 | | | | | | 3 | |
| | **C** | 4 | | | 4 | | 1 | | | | | | | | | | | |
| | **A** | 12 | | | 2 | | 10 | | | | | 4 | 1 | 13 | | | 11 | |
| | | **-8** | **-7** | **-6** | **-5** | **-4** | **-3** | **-2** | **-1** | **S/T** | **1** | **2** | **3** | **4** | **5** | **6** | **7** | **8** |
| | | Position number to the left of S/T | | | | | | | | | Position number to the right of S/T | | | | | | | |

(Center column for amino acid rows: NOT APPLICABLE; right block header: CENTER)

* Variables in the LPM of Table 11 are shaded in green.

*How do fit statistics like the KS statistic and the Brier Score vary for the selected LPM?*

The KS statistic is an alternative measure to the well-known Gini coefficient ([105-107] and [108]). The KS statistic for the LPM in Table 11 is 99.2%; and the corresponding Brier Score is 0.0094 (see Figure 6). To study the out of sample variation in KS statistic and Brier Score, we reestimate the coefficients of the LPM in Table 15 under 5-fold CV, 100 times. That is, the predictors of this LPM are held "constant" and only their coefficients are reestimated by pooling the data in 4 of the 5 randomly selected folds; and then calculating the KS statistic and Brier Score on the remaining fold. This exercise is done 5 times for a random selection of 5 folds. That is, 5 LPMs are estimated by partitioning the 5 folds into 5 sets of "training" and "validation" samples; where a training sample includes the data in 4 of the 5 folds, and the remaining fold is used as the validation sample. In turn, random selection of 5 folds is done 20 times. Thus, 100 KS statistics and Brier Scores are produced on sample sizes of about 400 each (the estimation sample size is 2,070).

The LPM in Table 15 instead of the one in Table 11 is selected for this exercise because it is simpler for our purpose. In particular, the difference between the two LPMs is the correction for heteroscedasticity, which is only a problem impacting the standard error of the coefficients, not their consistency. The results of this exercise are shown in Figure 10, which indicates that the variation in these fit statistics is restricted to fairly narrow bands. Thus, there is nothing unreasonable in LPM performance.

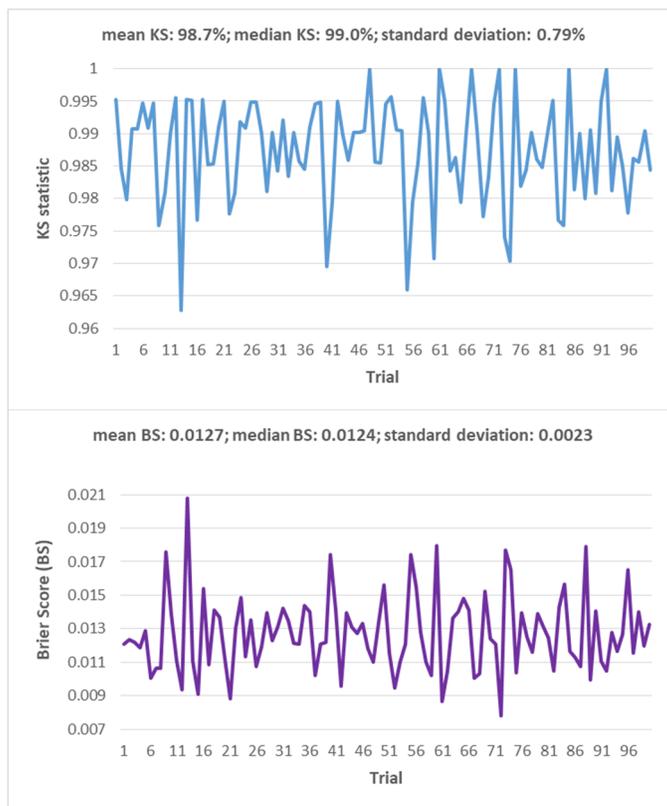

**Figure 10**: **KS statistic and Brier Score out-of-sample variations for the LPM**

## *Is "ASA_zero" a valid, rather than spurious, predictor in the LPM?*

*ASA* measures the extent to which a sequence is exposed. As mentioned previously there are two sequences in our data that have *ASA* values of 0, but are deemed to be *O*-glycosylated. In our model, *ASA_zero* is a statistically significant variable. Because of its special status as an "observation specific" dummy variable, it is not estimable within the classical logit modeling framework [71]. Its statistical significance, and persistence during CV, raises the open question of whether sequences that are not exposed at all have a greater propensity to be *O*-glycosylated. Although, we do not know the answer to that, we think that until empirical evidence to the contrary comes up, it does not hurt keeping *ASA_zero* in the LPM. Furthermore, the good news is that dropping it from the model does not change any of our major conclusions. That is, in our case, dropping *ASA_zero* as an explanatory variable, simply allowed its effect to be distributed among the other coefficients without disrupting anything. However, we did not drop it, because it is of interest to consider (until evidence to the contrary) its marginal contribution toward encouraging *O*-glycosylation.

The presence of *ASA_zero* in the LPM, but not *ASA*, raises another important question: is *ASA_zero* a spurious dummy variable in the LPM, because the fundamental variable underlying it, *ASA*, is not in the LPM? The answer is "no". And we discuss the reason for this next.

The ultimate assumption underlying this paper is that the amino acids and the positions they occupy along the sequence are the "building blocks" leading to *O*-glycosylation. This means a significant amount of the information carried by *ASA* should be embedded in the amino acids and the positions they occupy in the sequences. To test this assumption, we searched for a linear regression model with the natural logarithm (*log*) of *ASA* as the dependent variable and sequence information as independent variables. Our search followed the same principles that were used to model *psi*. We used the *log* transformation to get *ASA* "closer" to normality, and, as a consequence, the two observations with *ASA* values of 0 also get eliminated from the estimation dataset (because *log*(0) is undefined). *ASA*, like *psi*, has "heavy" tails. The selected model is shown in Table 23. The model has 103 predictors and an adjusted $R^2$ of about 80%. Its in-sample average squared error (ASE) is about 0.0425. The studentized residual distribution in Figure 11 show the heavy tails; but, a reasonable amount of symmetry is preserved. The model predicts that the two sequences with Uniprot Accession Numbers P02730 and P 32119, which have empirical *ASA* values of 0 (see Table 13), have predicted *ASA* values of about 32.5 and 24.0, respectively.

The information content in *ASA* generated by the amino acids and the positions they occupy along the sequence can be measured by how well the regression model estimates of *log(ASA)* can discriminate between *O*- and non-*O*- glycosylated sequences (i.e., *Y=1* vs. *Y=0*). The KS statistic would be a measure of this. In the estimation sample, if *Y* is ranked by estimated *log(ASA)*, the resultant KS statistic is about 32.0%, which is a strong enough value to not reject the assumption that the distribution of amino acids along the sequence does meaningfully drive *ASA*. Like in the case of *psi*, we do 5-fold CV (holding the 103 predictors "constant"), 20 times, to study the variability in the resultant 100 realizations of the out of sample ASEs and KS statistics. These variabilities, which are not unreasonable relative to the baseline ASE and KS statistic, are shown in Figure 12. The baseline ASE and KS statistic are for the model in Table 23, which is fit on the full sample.

**Table 23: Regressing *log(ASA)* against sequence information**

| Variable | $\beta$ | White's $|t|$ | Variable | $\beta$ | White's $|t|$ | Variable | $\beta$ | White's $|t|$ |
|---|---|---|---|---|---|---|---|---|
| intercept | 3.8265 | 79.93 | m6G | 0.2437 | 3.04 | p4A | -0.1698 | 2.96 |
| m1C | 0.4005 | 4.78 | m6K | 0.2509 | 4.17 | p4C | -0.4837 | 2.44 |
| m1E | -0.1849 | 2.76 | m6M | -0.2652 | 2.11 | p4D | -0.4424 | 6.50 |
| m1F | -0.3802 | 9.02 | m6W | 0.2150 | 2.77 | p4F | -0.2666 | 6.08 |
| m1G | -0.1734 | 4.07 | m7I | -0.2036 | 3.95 | p4G | -0.5605 | 8.13 |
| m1I | -0.4253 | 7.95 | m7K | 0.4807 | 7.38 | p4H | -0.2247 | 3.50 |
| m1K | 0.2688 | 3.63 | m7P | 0.1391 | 2.64 | p4I | 0.3184 | 7.64 |
| m1M | -0.1687 | 2.12 | m7Q | 0.3848 | 5.22 | p4S | -0.1250 | 2.75 |
| m1Q | 0.4901 | 6.87 | m7R | -0.2030 | 2.36 | p4W | -0.1784 | 2.80 |
| m1S | 0.1580 | 2.89 | m7S | 0.2185 | 3.14 | p5A | 0.1559 | 3.37 |
| m1V | 0.0999 | 2.51 | m7V | 0.2153 | 4.68 | p5C | -0.1432 | 2.21 |
| m1Y | -0.2066 | 3.66 | m7W | 0.2831 | 2.64 | p5D | -0.1959 | 3.46 |
| m3F | -0.4134 | 7.04 | m8G | -0.5119 | 11.04 | p5K | 0.3154 | 6.66 |
| m3G | 0.1102 | 2.47 | m8S | -0.1997 | 4.90 | p5V | 0.2375 | 4.07 |
| m3I | -0.4745 | 6.72 | m8Y | 0.6461 | 11.36 | p5W | 0.3914 | 3.95 |
| m3L | -0.2031 | 4.83 | p1C | 0.3582 | 5.47 | p6A | 0.1402 | 3.83 |
| m3M | -0.2393 | 3.21 | p1K | 0.3864 | 5.90 | p6E | -0.2923 | 6.89 |
| m3S | 0.1590 | 2.64 | p1L | 0.1875 | 4.96 | p6H | 0.1935 | 3.36 |
| m3V | -0.2821 | 4.05 | p1Q | -0.1655 | 2.41 | p6I | 0.3046 | 4.23 |
| m3W | -0.2581 | 4.16 | p1V | 0.1280 | 3.00 | p6K | -0.1276 | 1.92 |
| m3Y | -0.1855 | 3.34 | p2E | 0.1752 | 3.76 | p7C | 0.1051 | 1.73 |
| m4C | 0.4717 | 3.35 | p2F | 0.1018 | 1.72 | p7H | 0.2926 | 4.07 |
| m4D | 0.3255 | 3.79 | p2G | -0.2754 | 4.66 | p7I | -0.2492 | 3.18 |
| m4F | 0.3545 | 7.04 | p2K | -0.1480 | 2.10 | p7P | 0.1066 | 2.49 |
| m4K | 0.1518 | 2.83 | p2N | -0.1830 | 2.88 | p7Q | -0.1690 | 2.20 |
| m4L | 0.1450 | 3.83 | p2V | -0.1858 | 3.33 | p7T | 0.0952 | 1.96 |
| m4P | 0.4735 | 8.35 | p3A | -0.2814 | 5.72 | p7W | -0.2539 | 2.93 |
| m4Q | 0.3059 | 5.66 | p3C | 0.2717 | 3.90 | p7Y | 0.2480 | 3.36 |
| m4S | 0.1391 | 2.63 | p3D | -0.2040 | 4.27 | p8G | 0.1935 | 3.69 |
| m4T | -0.1328 | 2.2 | p3H | -0.2118 | 2.19 | p8H | 0.2042 | 3.94 |
| m5I | -0.4287 | 5.15 | p3I | -0.1469 | 3.65 | p8K | 0.3261 | 7.20 |
| m5P | 0.1189 | 1.72 | p3K | -0.4302 | 5.66 | p8L | 0.1804 | 2.63 |
| m5R | -0.2278 | 3.95 | p3L | -0.2711 | 6.92 | p8R | 0.2939 | 5.12 |
| m5Y | -0.3245 | 5.27 | p3R | -0.1781 | 2.87 | pos | 0.1555 | 2.65 |
| m6E | 0.1342 | 2.47 | p3Y | -0.2788 | 2.90 | | | |

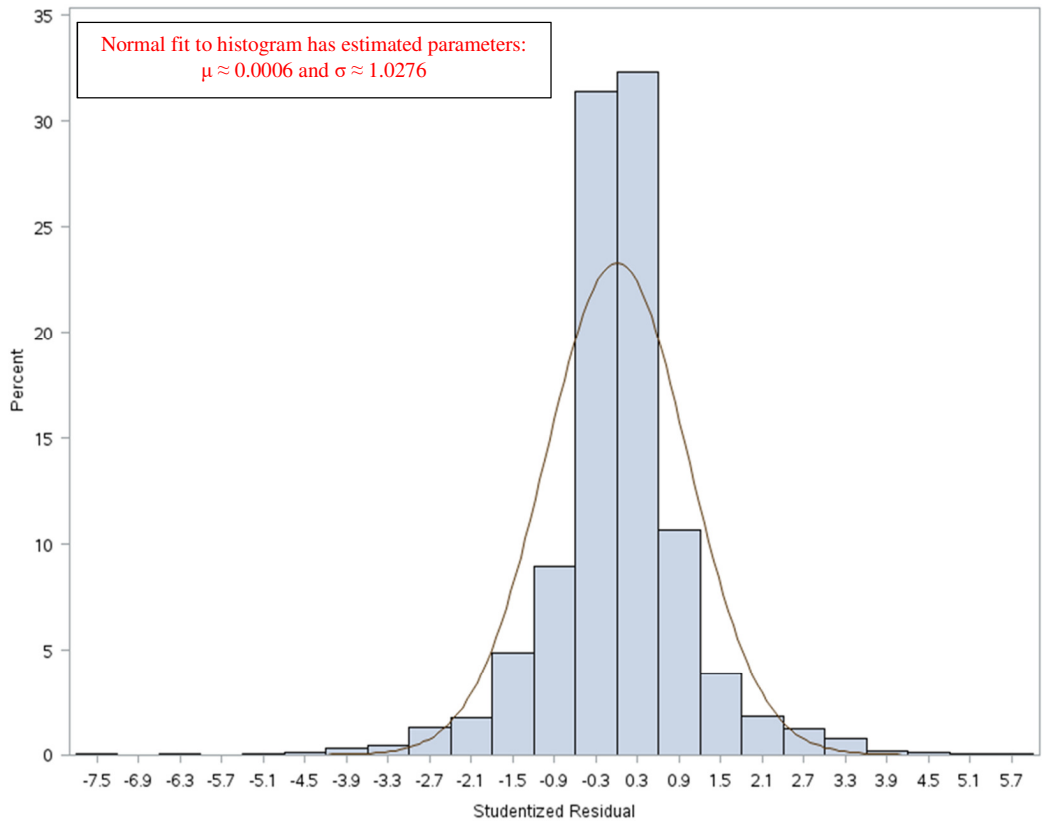

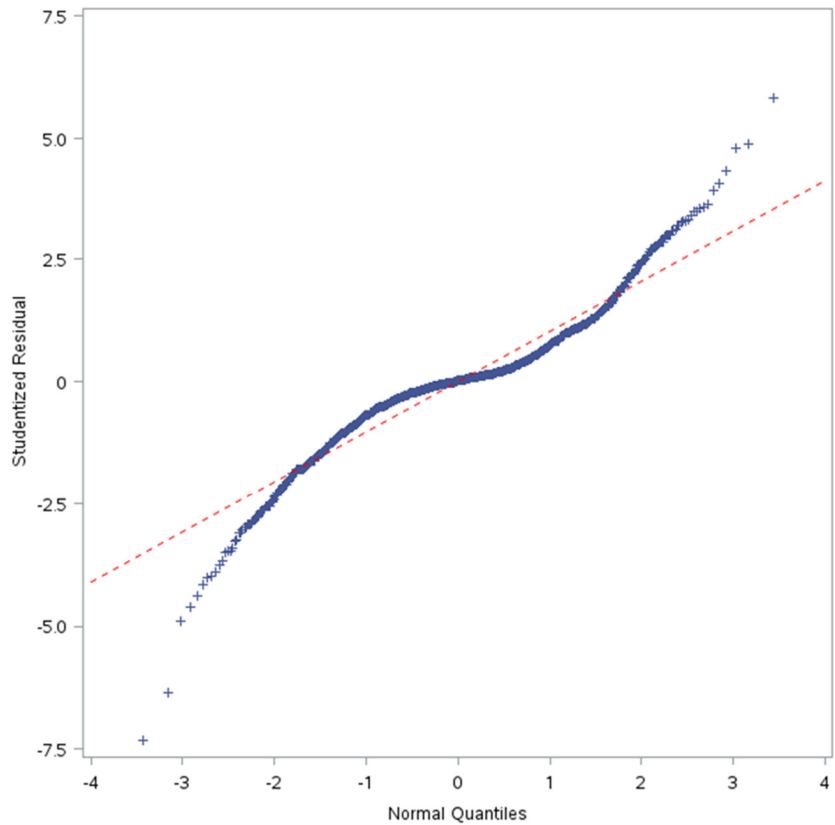

**Figure 11**: Distribution of studentized residuals for the model in **Table 23**

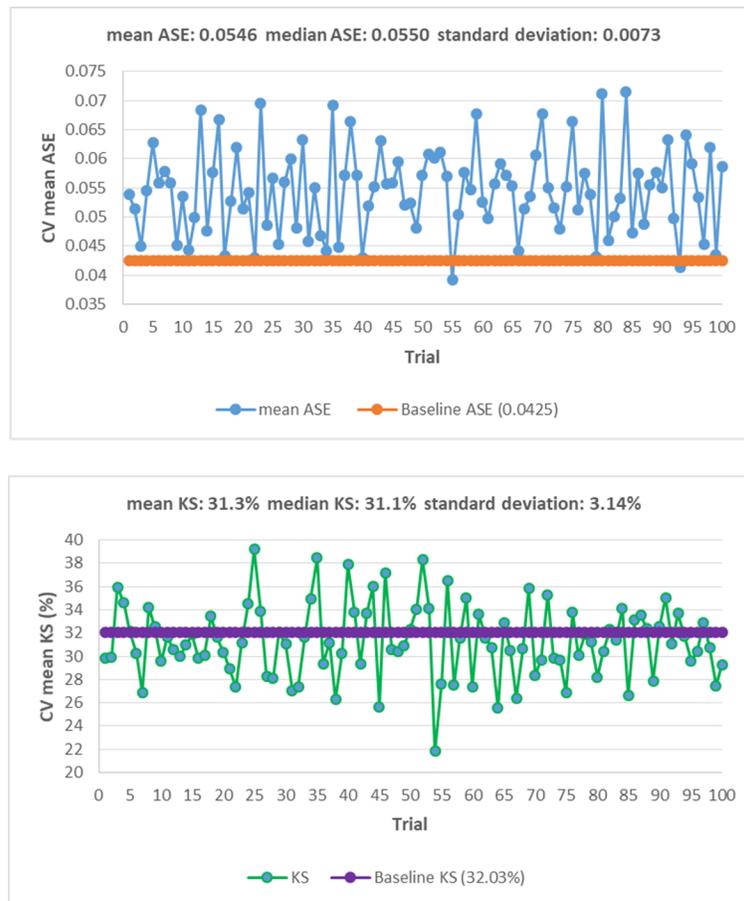

**Figure 12**: **Variations in ASE and KS over 100 CV trials generated by the model in Table 23**

### *Is it valid to use mixed data from PTMs occurring in different parts of the cell to estimate the LPM?*

The PTMs occur in different parts of the cell by a very complex process involving many different enzymes. For example, the initial processing of the *N*-glycosylation occurs in the rough ER and the process ends in the Golgi apparatus; phosphorylation occurs in the cytoplasm/nucleus; and *O*-GlcNAc proteins are also cytosolic/nuclear. So, the question is whether our "control" dataset (i.e., *N*-glycosylated sequences) is the correct "negative control"? The answer is "yes". By virtue of their differences in their mechanisms and cellular compartmentalization, it makes logical sense to use *N*-glycosylated sequences as the negative control for *O*-glycosylation. For example, in the absence of a consensus sequon for *O*-glycosylated sequences, using sequences with the *S/T* motif within the *O*-glycosylated set of sequences makes little sense because there is no guarantee that such sequences will not get *O*-glycosylated.

The LPM models a binary outcome. For example, when tossing a "fair" coin, if one wants to model (using the LPM, say) whether the coin will come down as "head", then one needs to start with two distinct data subsets: a set of heads and a set of tails [109]. One cannot have entries in the set of tails (the "control") that may also be heads. From physics, we know that the outcome of the toss is exact, not probabilistic [109]. However, given the mathematical complexity of, and the instrumentation needed for, making this prediction based on the laws of physics (e.g., Newton's laws), we use the language of probability, instead of physics, to estimate the outcome [110]. In particular, we say the probability of "head" is 50%. But, although this statement is false in theory, the methodology used is an important approximation that can be quite useful in practice, and for guiding experimental work.

Similarly, our use of the complex concept called "probability" in this paper does not imply that we are saying there is a fundamental or intrinsic randomness in how a sequence folds and becomes sugar linked; rather, we are saying that given the underlying complexity of this process (as in the case of the coin toss), our methodology, which depends on probabilistic ideas, may represent a reasonable approximation (via rank-ordering) to facilitate predicting whether a sequence is *O*-glycosylated [85]. Now the way we do this, is by starting with "partial" or "incomplete" information. That is, we start with explanatory variables in the ±10 positions around the *S/T* site. The distribution of amino acids in this neighborhood is considered. And the structural information at *S/T* is considered. For example, we have not considered the distribution of amino acids further away from *S/T*, which must also carry key information. So, for example, the LPM is "unaware" of the existence of the signal peptide in *N*-glycosylated proteins that sends it to a specific part of the cell.

So, we assumed that sequence and structural information in the neighborhood of the *S/T* site is sufficient for predicting the likelihood of *O*-glycosylation. To validate this assumption, we started with Table 9, which indicates that the proportion of *O*-glycosylated proteins having the $N - \sim P - S/T$ sequon is tiny (1.22% of *O*-GlcNAc and 0.77% of *O*-GalNAc proteins). Given this empirical observation, we assumed that the "signature" $N - \sim P - S/T$ discourages *O*-glycosylation – i.e., it appears to be an approximate inhibitor of *O*-glycosylation. Now, a reason that only a tiny number of sequences with the $N - \sim P - S/T$ sequon are *O*-glycosylated may be because the remaining signature around the *O*-glycosylated *S/T* site is quite similar to the corresponding one around the *N*-glycosylated *S/T* site. And, there may be some "signal" outside of our chosen ±10 neighborhood that may occasionally allow a protein with the $N - \sim P - S/T$ sequon to get *O*-glycosylated. However, the LPM is not trained on data outside of the ±10 neighborhood and, thus, such a signal, if it were to exist, is unobservable from the LPM's frame of reference. So, our selecting the *N*-glycosylated sequences as the negative control set is independent of cell "compartments". We are simply training the LPM on the signatures around the ±10 neighborhood of the *N*- and *O*- glycosylated sequences. Furthermore, if our assumption regarding the signature is unreasonable, we would have seen material deterioration in prediction accuracy under CV, but did not.

The consensus *composite* sequon for *O*-glycosylation is ~ *(W–S/T–W)*, where "~" denotes the "not" operator. This means *~W – S/T – W*, *W – S/T – ~W*, or *~W – S/T – ~W* are necessary for *O*-glycosylation. The RR estimated LPM approach was instrumental in making this finding. The consensus sequon for phosphorylation is ~ *(W–S/T/Y/H–W);* although *W–S/T/Y/H–W* is not an absolute inhibitor of phosphorylation. Structural attributes (beta turn II, II´, helix, beta bridges, beta hairpin, and the phi angle) are significant predictors of *O*-GlcNAc glycosylation.

For LPM estimation it is found that *N*-glycosylated sequences are good approximations to non-*O*-glycosylatable sequences; although *N – ~P – S/T* is not an absolute inhibitor of *O*-glycosylation. The selective positioning of an amino acid along the sequence, differentiates the PTMs of proteins. Some *N*-glycosylated sequences are also phosphorylated at the *S/T*-site in the *N – ~P – S/T* sequon. ASA values for *N*-glycosylated sequences are stochastically larger than those for *O*-GlcNAc glycosylated sequences.

The LPM with sequence *and* structural data as explanatory variables yields a Kolmogorov-Smirnov (KS) statistic value of 99%. With only sequence data, the KS statistic erodes to 80%. With 50% as the cutoff probability for predicting *O*-GlcNAc glycosylation, this LPM mispredicts 21% of out-of-sample *O*-GlcNAc glycosylated sequences as not being glycosylated. The 95% confidence interval around this mispredictions rate is 16% to 26%.

Several other studies are available for the predicting the likelihood of *O*-glycosylation. These studies have employed "machine learning" techniques, like random forests and factor analysis, to predict *O*-glycosylation [111-114]. However, this approach was not pursued in this paper because the focus herein is on the *explanatory process* driving *O*-glycosylation. Thus, classical hypothesis testing is germane to the present work, and algorithmic modeling with its pervasive focus on prediction is not, in this paper, an end in itself. In this sense, Cox's crucial comment to Breiman is echoed, which is that the "starting point" in this paper is not the data, but the underlying process generating it [115]. Furthermore, echoing Cox, again, the preference in this paper is to avoid proceeding with "a directly empirical black-box approach" in favor of trying to "take account of some underlying explanatory process" that can be simply represented by the LPM. In this context, the Lasso penalized logistic model was also fit to the data. However, it was found that the LPM is quite competitive with it, and provides a simpler mechanistic understanding of the *O*-glycosylation prediction process.